\documentclass[useAMS,usenatbib]{mn2e}
\usepackage{color}
\usepackage{epsfig,amsmath,natbib}

\usepackage{color}
\usepackage{ulem} 

\newcommand{\BellAmp}{N_B} 
\newcommand{\Eover}{\overline{\mathcal{E}}}
\newcommand{\Delfiso}{\delta f_\mathrm{iso}}
\newcommand{\vkvec}{\mathbf{v}(\mathbf{k})}
\newcommand{\bkvec}{\mathbf{b}(\mathbf{k})}
\newcommand{\hydro}{hydrodyamical}
\newcommand{\epCR}{\epsilon_\mathrm{cr}}
\newcommand{\ECR}{E_\mathrm{cr}}
\newcommand{\Tmax}{\theta_\mathrm{max}}
\newcommand{\Vmean}{\overline{\mathbf{V}}}
\newcommand{\PCR}{P_\mathrm{cr}}
\newcommand{\fCR}{f(\mathbf{r},\mathbf{p},t)}
\newcommand{\Ptensor}{\Pi^{(r)}_{\alpha\beta}}

\newcommand{\PrVec}{\mathbf{P}^{(r)}}
\newcommand{\kapt}{\kappa_t}
\newcommand{\bSqMean}{\langle \mathbf{b}^2\rangle}
\newcommand{\taucor}{\tau_\mathrm{cor}}
\newcommand{\alpt}{\alpha_t}
\newcommand{\delb}{\delta \mathbf{b}}
\newcommand{\jzCR}{j_0^\mathrm{cr}}
\newcommand{\Bbar}{\overline{\mathbf{B}}}
\newcommand{\Vbar}{\overline{\mathbf{V}}}
\newcommand{\dVbar}{\delta\overline{\mathbf{V}}}
\newcommand{\Bover}{\overline{B}}
\newcommand{\jCRmean}{\overline{j^\mathrm{cr}}}
\newcommand{\jCRmeanVec}{\overline{\mathbf{j}^\mathrm{cr}}}
\newcommand{\jCRVec}{\mathbf{j}^\mathrm{cr}}
\newcommand{\jCRx}{\overline{j^\mathrm{cr}_{x'}}}
\newcommand{\jCRy}{\overline{j^\mathrm{cr}_{y'}}}
\newcommand{\jCRz}{\overline{j^\mathrm{cr}_{z'}}}
\newcommand{\djCRx}{\delta j^\mathrm{cr}_x}
\newcommand{\djCRy}{\delta j^\mathrm{cr}_y}

\newcommand{\GamMax}{\gamma_\mathrm{max}}
\newcommand{\kpar}{k_{\parallel}}
\newcommand{\Bpar}{B_{\parallel}}
\newcommand{\jpar}{\mathbf{j}_{\parallel}}
\newcommand{\rgz}{r_{g0}}
\newcommand{\JCR}{\mathbf{J}^\mathrm{cr}}
\newcommand{\jCR}{\mathbf{j}^\mathrm{cr}}
\newcommand{\nCR}{n_\mathrm{cr}}

\newcommand{\SCly}{self-consistently}
\newcommand{\Lwave}{long-wavelength}

\newcommand{\CRh}{cosmic-ray}
\newcommand{\syn}{synchrotron}
\newcommand{\synch}{synchrotron}
\newcommand{\nonrel}{nonrelativistic}
\newcommand{\rel}{relativistic}
\def\lsim{\;\raise0.3ex\hbox{$<$\kern-0.75em\raise-1.1ex\hbox{$\sim$}}\;}
\def\gsim{\;\raise0.3ex\hbox{$>$\kern-0.75em\raise-1.1ex\hbox{$\sim$}}\;}

\def\alf{Alfv\'en }

\def\lsim{\;\raise0.3ex\hbox{$<$\kern-0.75em\raise-1.1ex\hbox{$\sim$}}\;}
\def\gsim{\;\raise0.3ex\hbox{$>$\kern-0.75em\raise-1.1ex\hbox{$\sim$}}\;}

\title[Cosmic-ray current driven turbulence in shocks]{Cosmic-ray current driven turbulence in shocks with efficient
particle acceleration: the oblique, long-wavelength mode
instability}
\author[A.M.Bykov, S.M.Osipov, D.C.Ellison]{A.M.Bykov$^{1}$\thanks{E-mail:byk@astro.ioffe.ru}, S.M.Osipov$^{1}$, D.C.Ellison$^{2}$\\
$^{1}$Ioffe Institute for Physics and Technology, 194021
St.Petersburg, Russia\\
$^{2}$Physics Department, North Carolina State University, Box 8202,
Raleigh, NC 27695}

\begin{document}


\pagerange{\pageref{firstpage}--\pageref{lastpage}} \pubyear{}

\maketitle

\label{firstpage}

\begin{abstract}
In order for diffusive shock acceleration (DSA) to accelerate
particles to high energies, the energetic particles must be able to
interact with magnetic turbulence over a broad wavelength range. The
weakly anisotropic distribution of accelerated particles, i.e.,
cosmic rays (CRs), is believed capable of producing this turbulence
in a symbiotic relationship where the magnetic turbulence required
to accelerate the CRs is created by the accelerated CRs themselves.
In efficient DSA, this wave-particle interaction can be strongly
nonlinear where CRs modify the plasma flow and the specific
mechanisms of magnetic field amplification.
Resonant interactions have long been known to amplify magnetic
fluctuations on the scale of the CR gyroradius,  and
\citet{bell04} showed that the CR current can efficiently amplify
magnetic fluctuations with scales smaller than the CR gyroradius.
Here, we show with a multi-scale, quasi-linear analysis that the
  presence of turbulence with scales shorter than the CR gyroradius
  enhances the growth of modes with scales longer than the
  gyroradius, at least for particular polarizations.  We
  use a mean-field approach to average the equation of motion and the
  induction equation over the ensemble of magnetic field oscillations
  accounting for the anisotropy of relativistic particles on the
  background plasma.
We derive the response of the magnetized CR current on magnetic
field fluctuations and show that, in the presence of short-scale,
Bell-type turbulence, long wavelength modes are amplified.
The polarization, helicity, and angular dependence of the growth
rates are calculated for obliquely propagating modes for wavelengths
both below and above the CR mean free path.  The long-wavelength
growth rates we estimate for typical supernova remnant parameters
are sufficiently fast to suggest a fundamental increase in the
maximum CR energy a given shock can produce.
\end{abstract}

\section{Introduction}
Diffusive shock acceleration (DSA) is the most promising mechanism for
  producing superthermal and \rel\ particles in a wide range of
  astrophysical objects ranging from the Earth bow shock to shocks in
  galaxy clusters \citep{be87,je91,md01}. While this mechanism is
  believed to be efficient and capable of producing cosmic rays (CRs) to
  energies well above $10^{15}$\,eV in supernova remnants (SNRs), fast
  and efficient DSA demands that particles interact strongly with large
  amplitude magnetic fluctuations in the shock vicinity.  The amplitude
  of the required MHD turbulence is substantially higher than the
  ambient MHD turbulence forcing a bootstrap scenario where the
  accelerated particles generate the turbulence required for their
  acceleration.  Direct support for the modest self-generation of MHD
  turbulence has long been seen in heliospheric shocks
  \citep[e.g.,][]{Gurnett1985}
and more recent observations of
  X-ray \synch\ radiation from several young SNRs provide indirect
  evidence for extreme super-adiabatic magnetic field amplification
  (MFA) associated with CR production and DSA \citep[see
  e.g.,][]{vl03,bambaea05,uchiyamaea07,vink08}.

A critical aspect of MFA concerns the dynamic range of the
self-generated turbulence.  Since the maximum CR energy a given shock
can produce is determined largely by the power in the longest wavelength
turbulence -- that turbulence which is needed to trap CRs with the
largest gyroradii -- the production of long-wavelength turbulence must
be included in any full description of nonlinear DSA.
The essential features of our calculation for the growth rate of
turbulence with scales larger than the gyroradii of the generating
CRs as these CRs interact with turbulence with scales shorter
than their gyroradii, were presented for parallel propagating
shocks in \citet{bot09}.  Here, a more complete calculation is
given which includes oblique shock geometry.

The situation where CRs interact with short-wavelength turbulence is
just what is expected in a shock precursor where the CR current
efficiently generates purely growing modes with wavelengths much shorter
than the gyroradii of the CRs \citep[i.e.,][]{bell04}.
For wavelengths both below and above
the CR mean free path, we perform a multi-scale, linear
calculation that takes into account the polarization, helicity, and
angular dependence of the growth rates for obliquely propagating modes.
An important question for DSA and CR origin has always centered around
the maximum particle energy a given shock can produce. For a shock of a
given size, age, and magnetic field geometry, the maximum CR energy
depends totally on the power in the longest wavelength turbulence. Our
estimates suggest that the growth rates for long-wavelength modes are
fast enough to allow a significant increase in CR energy.

The study of turbulence generation associated with CRs and DSA has
a long history. Magneto-hydrodynamic (MHD) type wave amplification
due to the resonant cosmic-ray streaming instability was studied
in the context of galactic cosmic-ray origin and propagation since
the 1960s \citep[see e.g.][]{kc71,wentzel74,achterberg81,berea90,
zweibel03}. It was proposed by \citet{bell78} as a source of
magnetic turbulence in the test particle DSA scenario, and
nonlinear models of DSA including streaming instabilities and MFA
were investigated by \citet{ab06},\citet{veb06} and \citet{rkd09}.
A Monte Carlo model of nonlinear DSA with MFA from resonant
instabilities induced by accelerated particles, which also
accounted for the effects of dissipation of turbulence upstream of
a shock and the subsequent precursor plasma heating, was developed
by \citet{vbe08}.  The Monte Carlo work showed that strong
feedback effects between the plasma heating due to turbulence
dissipation and particle injection are important for understanding
the nonlinear nature of efficient DSA. While the resonant
instability is arguably the simplest instability thus far studied,
a full description is still beyond any single technique either
analytic or computer simulation. Mixed techniques are required
where analytical recipes to account for MFA, dissipation, and
other effects are blended with simulations.

In addition to resonant instabilities, a number of non-resonant
instabilities have been investigated for DSA.  A non-resonant acoustic
instability, where the \CRh\ pressure gradient in the shock precursor
amplifes compressional disturbances, was investigated by \citet{dd85}
and \cite{df86}.  \citet{berezhko86} and \citet{chalov88} generalized this by
accounting for a regular magnetic field.  The obliquely propagating
magnetosonic modes in the inhomogeneous precursors of \CRh\ modified
shocks were further examined by \citet{zankea90} who also
investigated the role of strong, intermediate, and weak \CRh\ scattering
regimes. The effects of the acoustic instability on the particle
distribution were investigated by \citet{kangea92} using a
time-dependent numerical simulations of the diffusion-advection
transport equations. The non-linear evolution of unstable acoustic waves
in the precursors of \CRh\ mediated shocks with large Mach numbers
was shown by \citet{chalov10} to result in the formation of a host of
small, weak shock waves that could heat the thermal plasma and,
therefore, change the parameters of the strong, large-scale shock.
%
Other nonlinear work was done by \citet{md09}
where they suggested that the development of unstable acoustic waves
might result in shock-train formation that could speed up the
acceleration rate and stimulate the inverse cascade of \alf waves
generated by the accelerated particles.  Recently, \citet{beresnyakea09}
proposed a model where the stochastic magnetic fields in the shock
precursor are generated through a short-scale dynamo mechanism. In this
model, the fluid velocity turbulence driving the dynamo is produced
through interactions of the CR pressure gradient and density
perturbations in the precursor.

A large amount of recent work has been stimulated by Bell's discovery of
a fast, non-resonant instability driven by the CR current
\citep[][]{bell04}.  In this instability, the CR current in the shock
precursor drives purely growing, incompressible, electromagnetic
modes. These modes have wavevectors along the magnetic field and have
wavelengths much shorter than the CR gyroradius
radius. Because of the short wavelengths, the CR current is essentially
unmagnetized in this regime and induces a return current in the
background plasma which is nearly parallel to the locally homogeneous
magnetic field. This allows the approximation that the current
generating CRs are only weakly perturbed by the magnetic fluctuations
they create (we shall discuss this instability in some detail in section
\ref{SSD}).

The focus of much of the analytical and numerical work on the Bell
instability concerns the saturation level and the spectral properties of
the instability \citep[see
e.g.][]{pelletierea06,marcowithea06,ab09,lm09,vbe09,ze10}.  If the
approximation that the CR current is only weakly affected by the
short-scale fluctuations holds, nonlinear MHD-type simulations, where
the CR current is fixed as an external parameter, can be used
\citep[e.g.][]{bell04,zp08,zpv08,rsdk08}. These simulations typically
show a rather high saturation level of the self-generated, short-scale
turbulence and also show nonlinear spectral energy transfer (i.e., the
cascading of turbulence energy) into both larger and smaller
scales. However, since long-wavelength magnetic fluctuations on scales
comparable to or larger than the CR gyroradius cause a non-negligible
response of the CR current, the MHD simulations are limited to
short-scale fluctuations only.
In a real system, long-wavelength fluctuations will occur and they
  will induce current perturbations perpendicular to the local
  mean magnetic field even in the linear regime
  \citep[see][]{bt05,bot09}. This important physical effect will be
  difficult for MHD simulations to capture.

Particle-in-cell (PIC) and hybrid plasma simulations\footnote{In this
  context, hybrid means protons are treated as particles and electrons
  are modeled as a charge-neutralizing background fluid.}
are widely used with different CR and background plasma
parameters. They can explore both linear and nonlinear regimes of
the non-resonant mode, the saturation level, and  spectral energy
transfer mechanisms
\citep{niemiecea08,rs09,ohiraea09,stromanea09,rs09a,gargatea10}.
In the linear regime, the simulated growth rates confirm the basic
theoretical predictions of \citet{bell04}. Amplification factors
above 10 for the magnetic field were achieved in some simulations.
Plasma acceleration along the drift motion of CRs saturates the
instability at the magnetic field level when the \alf velocity in
the amplified field approaches the drift velocity of CRs. The
instability can also saturate earlier if CRs get magnetized by the
amplified field. If the CR current is transverse to the initial
magnetic field then both hydrodynamic \citep[e.g.,][]{bell05} and
PIC simulations by \citet{rs09a} have shown the presence of a
fast, current driven instability.

Magnetic field amplification in the long-wavelength regime, where
the gyroradii of the ``magnetized'' CRs are smaller than the
wavelength of the magnetic perturbations, was discussed by
\citet{bot09}. This work used a quasi-linear kinetic equation
approach for modes parallel to the initial magnetic field.
It was demonstrated by \citet{bot09} that the \Lwave\
instability, produced by the perturbed, perpendicular CR current,
was strongly affected by the short-scale turbulence produced by the
Bell instability.  This coupling of long and short-scale
instabilities, which is certain to be nonlinear if DSA is efficient,
places significant demands on plasma particle and fluid code
simulations since they must cover a wide dynamic range to fully
describe the problem from injection to maximum CR energy. These
demands are particularly severe for \nonrel\ shocks with parameters
typical of SNRs \citep[see][]{vbe08}.

In the present paper we study the properties of long-wavelength, oblique
modes in DSA for different assumptions on the CR scattering rate in
background, short-scale turbulence.  The long-wavelength growth rates we
calculate depend on the CR current interacting with the short-wavelength
turbulence generated by Bell's instability, which has the fastest growth
rate in the short wavelength regime and produces the mode polarizations
we assume.
While our calculation is linear in $\Delta B/B$, any application in a
realistic DSA scenario would assume efficient acceleration with a
non-negligible fraction of shock ram pressure being transformed to CRs.
Our estimates given in Section~\ref{Disc} for young SNRs, suggest that,
for some parameters at least, the production of long-wavelength
turbulence in the forward shock precursors of young SNRs
will be important.
Since the shock structure depends critically on the
efficiency of turbulence amplification, cascading, and dissipation, our
work must be viewed as a step towards the development of a nonlinear
model of DSA where all of the following are treated \SCly:
(1) efficient particle injection and acceleration occur;
(2) particles of different energies participate differently in the
instability generation and magnetic field amplification;
(3)
turbulence cascading and dissipation are accounted for; and
(4) the nonlinear feedback of particles and fields on the bulk flow is
    included \SCly.
To our knowledge, the only techniques that are currently capable of
including all of these coupled effects in a calculation suitable for
modeling DSA in SNRs, even in parameterized form, are the semi-analytic
methods of Blasi and co-workers \citep[see][ and references
therein]{BAC2007,CBAV2008}, and the Monte Carlo methods of Vladimirov
and co-workers \citep[see][ and references therein]{vbe09}.
Before providing the details of our derivation we outline the steps in
the next Section.

\section{Outline of the Model}

Our goal is to obtain the growth rates for the cosmic-ray current
driven, long-wave instability in a shock precursor containing
short-scale fluctuations. The terms ``long'' and ``short'' are relative
to the minimum CR gyroradius\footnote{Due to the strong energy
  dependence of the CR diffusion coefficient, the CR distribution in a shock
precursor typically shows a fairly sharp cutoff at low energies. This is
the minimum energy we refer to and this
cutoff energy increases with distance upstream from the viscous
subshock.}
and the first step in our derivation is to obtain the equations of the
plasma and magnetic field dynamics averaged over the short-scale
fluctuations.

In Section~\ref{SSD} we describe the properties of the short-scale
CR-current driven modes produced by Bell's instability
\citep[][]{bell04,bell05}. We discuss the polarization properties of the
modes, the helicity they induce in the background plasma, and emphasize
how the mode polarization is important for the long-wavelength dynamics.

The equation of motion and the magnetic induction equation, both for
the background plasma and both averaged over the short-scale modes
that are needed to study the long-wavelength growth rates, are
discussed in Section \ref{meandyn}.
The detailed derivations of these averaged equations are given in the
Appendix. In Section \ref{MFE} of the Appendix we obtain the
averaged magnetic induction equation. In Section \ref{EMotion} we
average the momentum equation.
We generalized the mean field method developed in  dynamo
theory \citep[see][]{bf02, bs05, brandenburg09} to average the equations
accounting for the specific effects of the
CRs.  The distinctive feature of the derivation is that the CR current,
which is responsible for the instability, is fully accounted for.
The CR current produces specific polarizations of the modes and this
influences the ratio of their kinetic and magnetic energy densities.
This, in turn, modifies the mode correlation functions that
determine the turbulent transport coefficients in the mean field
approach. The correlation functions are presented in Section~\ref{corr}
of the Appendix.

To get the linear dispersion relation for the long-wavelength
fluctuations, one needs to know the response of the CR current to the
short-scale fluctuating magnetic field, that is, to the Bell
turbulence. This is an important point. For solely the Bell
instability, the response of the CR current to the short-scale
fluctuations can be ignored. In the long-wavelength regime, however, the
CR current changes in response to the magnetic field perturbations
imposed on the plasma system and these changes need to be accounted for.
Here, we calculate the CR response using a kinetic equation for the CR
distribution function with a collision operator that describes the
CR scattering by magnetic fluctuations. The details of the CR current
response derivation are presented in Section \ref{CRCR} of the Appendix.

Finally, the growth rates we obtain for long-wavelength fluctuations in
plasma systems with a CR current are presented in
Section~\ref{Disp}. Two distinct long-wavelength regimes are
discussed. The first regime is where the growing modes, at some position
in the shock precursor, have wavelengths between the gyroradius and the
mean free path of the lowest energy CRs present at that
precursor position.
We call this regime ``intermediate''
and these modes are complementary to those seen in mean-field
dynamo theory except now the modes are modified by the presence of
the CR current. These long-wavelength modes are produced by the
anisotropic CR distribution in the presence of strong, small-scale
fluctuations.
The second regime is where the growing modes have wavelengths longer
than the CR mean free path. We refer to this regime as
``hydrodynamical.''
We note that in the case of very strong CR scattering, that is
when the Bohm limit is obtained and the CR mean free path equals
the gyroradius, the intermediate regime has no dynamic range and
only the \hydro\ regime produces the long-wavelength instability.

\section{Short-scale dynamics and Bell's instability}\label{SSD}
In MHD-type flows with cosmic rays imbedded in a background plasma, the
momentum equation of the background plasma, including the Lorentz force,
is given by
\begin{equation}\label{eqMotiontot0}
 \rho\left(\frac{\partial\mathbf{u}}{\partial
 t}+(\mathbf{u}\nabla)\mathbf{u}\right)
 =- \nabla P +
\frac{1}{c}(\mathbf{j}\times\mathbf{B})+e(n_{i}-n_{e})\mathbf{E}+
\nu\triangle\mathbf{u},
\end{equation}
where $\mathbf{u}$ and $\mathbf{j}$ are the bulk velocity and the
electric current of the background plasma, respectively.  The viscosity
$\nu$ is due to Coulomb collisions (or due to plasma oscillations on
scales much less than the CR gyroradii) and $\nu$ is typically small for
the effects discussed here.
We assume quasi-neutrality for the whole system consisting of
background plasma ions of number density $n_{i}$, electrons of
number density $n_{e}$, and cosmic rays of number density $\nCR$. For
simplicity we consider cosmic-ray protons only such that
$n_{i}+\nCR =n_{e}$.
Both the background electric current $\mathbf{j}$ and the electric
current of accelerated particles $\JCR$ are the sources of
magnetic fields in Maxwell's equations, where the displacement current
can be omitted for the slow MHD-type processes under
consideration. Furthermore, we
assume ideal plasma conductivity in Ohm's law:
\begin{eqnarray}\label{Jcrtotal}
{\rm rot}\mathbf{B} = \frac{4\pi}{c}(\mathbf{j} + \JCR),
\quad
%
%
\mathbf{E}=-\frac{1}{c}(\mathbf{u}\times\mathbf{B}).
\end{eqnarray}
Then the induction equation is given by
\begin{equation}\label{largeInd0}
\frac{\partial\mathbf{B}}{\partial t}=\nabla \times(\mathbf{u}\times
\mathbf{B})+\nu_{m}\triangle\mathbf{B} ,
\end{equation}
where $\nu_{m}$ is the magnetic diffusivity due to the Coulomb
collisions  or MHD plasma oscillations.

To study instabilities in flows with initially quasi-homogeneous
magnetic fields, the global current neutralization condition should
be fulfilled in Eq.~(\ref{Jcrtotal}). Substituting
Eq.~(\ref{Jcrtotal}) in Eq.~(\ref{eqMotiontot0}), we obtain the
momentum equation in the following form \citep{bell04}
\begin{eqnarray}\label{eqMotiontot1}
& & \rho\left(\frac{\partial\mathbf{u}}{\partial t}+
 (\mathbf{u}\nabla)\mathbf{u}\right)
 =- \nabla P+ \frac{1}{4\pi}(\nabla\times\mathbf{B})\times\mathbf{B}-
    \nonumber \\
  & &
-\frac{1}{c}(\JCR -e \nCR
\mathbf{u})\times\mathbf{B}+\nu\triangle\mathbf{u} \ .
\end{eqnarray}
In this equation, the electric current of accelerated particles $\JCR$
is an external current in the momentum equation of the background
plasma. The current is governed by sources of energetic particles and by
local electromagnetic fields. The current $\JCR$ initiates a
compensating return current in the background plasma. \citet{bell04,
bell05} discovered that the system is unstable against linear
perturbations that are $\propto exp(\gamma t+i\mathbf{k}\mathbf{r})$ and
the return current drives nearly purely growing modes. Here, $\gamma$ is
the linear growth rate.
The wavenumbers of growing modes must satisfy the condition $k r_{g0} >
$ 1, where $\rgz = c p_0/(e \Bpar)$
is the gyroradius of an accelerated particle of momentum $p_0$. All
particles with $p > p_0$ contribute to mode growth.

In a cold plasma with sound speed $a_{0}$, well below the
\alf velocity $v_{a}$, the linear growth rate obtained by \citet{bell05}
depends only on the wavevector projection $k_{z}$ on the local mean
magnetic field, i.e.,
\begin{equation}\label{DispU}
\gamma\approx \GamMax k_z/k ,
\end{equation}
where
\begin{equation}\label{DispU1}
\GamMax = v_{a}\sqrt{k_{1}|k|-k^{2}}
\end{equation}
is the growth rate for the modes propagating along the mean field and
\begin{equation}\label{k1}
k_1 = \frac{4\pi}{c} \frac{\jCRmean}{\Bover}
\ .
\end{equation}
Here, the bar means the CR current and magnetic field are averaged over
fluctuations with scales below $\rgz$.

According to the linear analysis of \citet{bell05}, the wavenumber
of a growing mode must satisfy the condition $\rgz^{-1}< k < k_{1}$.
Therefore, the instability growth rate is higher than the \alf
frequency $v_{a}k$. Note that this condition for mode growth $\rgz
k_1 > 1$, together with Eq.~(\ref{k1}), implies that the anisotropy
of the relativistic particle distribution, $\delta_\mathrm{cr}$,
exceeds the ratio of the mean magnetic field energy density to the
energetic particle energy density $E_\mathrm{cr}$. That is,
$\delta_\mathrm{cr}
> B^2/(4 \pi E_\mathrm{cr})$, where the CR current is given by
$\jCRmean \approx \delta_\mathrm{cr} e n_{cr} c$.

The polarization of the growing mode, given by
\begin{equation}\label{pol}
b_{x}=i\frac{k_{z}}{|k_{z}|}b_{y} ,
\end{equation}
 is important and we use the fact that the kinetic energy density in the
  growing mode dominates over the magnetic energy density to get
  simplified mean field equations.
The linear relation between the amplitude of the growing magnetic field
$\bkvec$ and the velocity $\vkvec$ of the perturbations
is given by
\begin{equation}\label{IndBV}
\gamma \bkvec  =iB_{0}k_{z} \vkvec .
\end{equation}
This yields
\begin{equation}\label{b2v2}
|\vkvec|^{2}\approx
\frac{1}{4\pi\rho}\frac{k_{1}}{|k_{z}|}|\bkvec|^{2},
\end{equation}
provided that the kinetic energy density in the growing mode dominates
over the magnetic energy density because $k_{1}> k_{z}$.  This is in
contrast to \alf modes where the energy densities are equal.  In
Section~\ref{corr} [namely, Eq.~(\ref{Correlvv})], the mode polarizations
are used to express the pair correlation functions for the fluctuating
fields.


We note here that the cosmic-ray current has
only a weak response to the short-scale fluctuations, while its
response to fluctuations with scales longer than $\rgz$ is much larger.
Therefore, the CR current variations can be neglected when averaging the
dynamic equations over fluctuations produced by the Bell instability and
we perform this averaging in the next Section.
For the long-wavelength fluctuations, the
current variations must be considered and this averaging is done in
Section A of the Appendix.

\section{The averaged equations of large-scale dynamics}\label{meandyn}
Here we average the momentum equation of the background
plasma and the induction equation over the short-scale fluctuations
produced by Bell's instability to obtain the mean field dynamics
equation.
Since the linear growth rate is fast for the Bell mode with the
wavevector along the local mean magnetic field, the bulk velocity of the
background plasma $\mathbf{u}(\mathbf{r},t)$ and
$\mathbf{B}(\mathbf{r},t)$ both experience rapid, incoherent
fluctuations on scales that are small compared to $r_{g0}$, the
gyroradius of a relativistic particle.
The unstable modes with wavevectors along the mean magnetic field are
nearly incompressible making the averaging of the mass continuity
equation straightforward.
The time and spatial scales separation relations are
defined by
$\JCR  = \jCRmeanVec + \jCR$,
$\mathbf{B} = \mathbf{\overline{B}} + \mathbf{b}$ and
$\mathbf{u} = \mathbf{\overline{V}} + \mathbf{v}$, where $\jCRmeanVec$,
$\mathbf{\overline{B}}$, and $\mathbf{\overline{V}}$ are the averaged
electric current of accelerated particles, the averaged magnetic field,
and the background plasma bulk velocity, while $\mathbf{v}$ and
$\mathbf{b}$ are the short-scale bulk velocity and magnetic field.
We can now average over the ensemble of short-scale fluctuations,
placing these averages in angular brackets, and obtain the averaged
momentum equation Eq.~(\ref{eqMotiontot1}):
\begin{eqnarray} \label{aveMomEq}
& & \frac{ \partial\mathbf{\overline{V}}}{\partial t } +
(\mathbf{\overline{V}}\nabla)\mathbf{\overline{V}} =
-\left\langle(\mathbf{v}\nabla)\mathbf{v}\right\rangle
   +\frac{1}{4\pi\rho}\left\langle(\nabla\times\mathbf{b})\times\mathbf{b}\right\rangle-
 \nonumber \\
  & &
  -\frac{1}{\rho}\nabla P-
      \frac{1}{c\,\rho}(
(\jCRmeanVec -e\, \nCR
\mathbf{\overline{V}})\times\mathbf{\overline{B}})+
    \nonumber \\
  & &
  +\frac{1}{4\pi\rho}((\nabla\times\mathbf{\overline{B}})\times\mathbf{\overline{B}}),
 \label{largeEqMotion}
\end{eqnarray}
and the averaged equation of magnetic induction
\begin{equation}\label{largeInd1}
\frac{\partial\mathbf{\overline{B}}}{\partial t}=c\nabla\times
\mathbf{\overline{\mathcal{E}}}+\nabla
\times(\mathbf{\overline{V}}\times
\mathbf{\overline{B}})+\nu_{m}\triangle\mathbf{\overline{B}}.
\end{equation}
Here $\mathbf{\overline{\mathcal{E}}}=\left\langle \mathbf{v}\times
\mathbf{b}\right\rangle$ is the average turbulent  electromotive
force.
The coordinate system is defined relative to the
unperturbed magnetic field $\mathbf{e_z} ={\mathbf{B_0}}/B_0$.

\section{The long-wavelength cosmic-ray current driven modes}\label{Disp}
In the presence of a cosmic-ray current, Bell's instability results in
the fast growth of short-scale modes with wavelengths shorter than the
gyroradius of the cosmic-ray particles.
However, as we showed above when we obtained the mean field dynamic
equations averaged over the ensemble of short-scale motions, the strong
short-scale turbulence influences the plasma dynamics on scales larger
than the CR gyroradii producing the turbulence.
Equations (\ref{aveMomEq}) and (\ref{largeInd1}) are designed to be
applied to the dynamics of modes with scales larger than $r_{g0}$, i.e.,
CR particles are magnetized on these scales.

In diffusive shock acceleration with strong scattering, the particle
mean free path, $\Lambda$, is often taken to be $\Lambda = \eta \rgz$,
where $\eta \geq 1$ and $\eta = 1$ is the Bohm limit.
If $\eta > 1$ there are two different regimes for the large-scale
dynamics. The first regime, discussed next, corresponds to $\eta^{-1} <
kr_{g0} < 1$, where the CR particles have gyroradii small compared to
the turbulence scales and can be considered magnetized. The second
regime
is for $k \rgz < \eta^{-1}$, where the particle gyroradii are
large compared to the turbulence wavelengths. These modes are driven by
the transverse current of anisotropic, magnetized cosmic rays and are
discussed in Section~\ref{Hydro}.

\subsection{Long-wavelength current driven
modes in the intermediate regime  ($\eta^{-1} < kr_{g0} <
1$)}\label{Dynamo}
%
The dynamic equations averaged over the short-scale fluctuations, that
we obtained in Sections~\ref{MFE} and \ref{EMotion}, can now be used to
study the long-wavelength modes.
We start with the linearized equation of motion
Eq.~(\ref{largeEqMotion0}) and the induction equation for the mean field
Eq.~(\ref{MediumInd1}). Denoting the small departures of physical values
from their unperturbed magnitudes by $\delta$, and performing the
analysis in the rest frame of the unperturbed upstream plasma so that
$\Vbar = \dVbar$, we obtain:
\begin{eqnarray}\label{mediumEqMotion0}
& & \frac{\partial\delta\mathbf{\overline{V}}}{\partial t}
=-\frac{1}{\rho}\nabla P
-\frac{1}{c\,\rho}(j^{cr}_{0}\mathbf{e}_{z}\times\mathbf{\delta b})-
\nonumber \\
& &
-\frac{1}{c\,\rho}(\delta \mathbf{j}^{cr}\times
B_{0}\mathbf{e}_{z})
+\frac{1}{4\pi\rho}((\nabla\times\delta\mathbf{b})\times
B_{0}\mathbf{e}_{z})+
\nonumber \\
& &
+\frac{\kappa_t}{c\rho}\left(\left[\delta
j^{cr}_{x}-g^{\,\prime}\delta
b_{x}\right]\mathbf{e}_{y}-\left[\delta
j^{cr}_{y}-g^{\,\prime}\delta b_{y}\right]\mathbf{e}_{x}\right),
\end{eqnarray}
\begin{eqnarray}\label{MediumInd2}
& & \frac{\partial\delta\mathbf{b}}{\partial
t}=\nabla\times(\delta\mathbf{\overline{V}}\times
B_{0}\mathbf{e}_{z})+\frac{2}{\rho
c}\alpha_{t}\nabla\times\overline{j}^{cr}_{z}\mathbf{e}_{z}+
\nonumber \\
& &
+ \frac{1}{2\rho c}\alpha_{t}\nabla\times\left(\delta
j^{cr}_{x}\mathbf{e}_{x}+\delta j^{cr}_{y}\mathbf{e}_{y}\right)+
\nonumber \\
& &
+\frac{3}{2\rho c}\alpha_{t}\nabla\times\left(g^{\,\prime}\delta
b_{x}\mathbf{e}_{x}+g^{\,\prime}\delta b_{y}\mathbf{e}_{y}\right),
\end{eqnarray}
where
$g' = \jzCR/B_0 = e \nCR v_s/B_0$, $v_s$ is the shock velocity, and
$\alpt = \bSqMean \taucor$.
In these linearized equations we have neglected terms $\sim e \nCR
\dVbar$ compared to $g' \delb$ which is justified if
$v_{\rm{ph}}/v_s \ll 1$ (where $v_{\rm{ph}}=\omega/k$ is the phase
velocity of the modes defined in Section~\ref{CRC11}).
The correlation time $\taucor$ is the relaxation time of triple
correlations (see Sections \ref{MFE} and \ref{EMotion} in Appendix, and
namely Eq.~\ref{ElectrMot11}) and is approximately the turnover time of
the Bell turbulence. The parameter $\taucor$ is an important parameter
in the mean field approach \citep[e.g.,][]{bs05}.

From Eqs.~(\ref{mediumEqMotion0},\ref{MediumInd2}), and using
Eq.~(\ref{deltaJcrNu1}), one obtains the dispersion relation in
the form:
\begin{eqnarray}\label{dispersMedium}
& &
\omega^{2}\mp\omega
ikk_{0}\frac{\alpha_{t}}{4\pi\rho}\left[\frac{1}{2}A(x_{0})+
\frac{3}{2}\right]-k^{2}v_{a}^{2}\pm
\nonumber \\
& &
 \pm kk_{0}v_{a}^{2}\left(1+\frac{
\kappa_t}{B_{0}}\right)\left[A(x_{0})-1\right]=0,
\end{eqnarray}
where $k_0 = 4 \pi g'/c$,
$v_a = B_0/ \sqrt{4 \pi \rho}$, the function
$A(x_{0})$ is defined by Eq.~(\ref{coeffANu}) in Section \ref{CRC11},
  and $\kapt$ is a turbulent transport coefficient in the averaged
  equation of motion and is defined in Eq.~(\ref{largeEqMotion0}). The
  $\pm$ signs give the modes with the opposite circular polarizations
  in the dispersion relation.

To estimate the coefficients in Eq.~(\ref{dispersMedium}) we assume
$\bSqMean \approx B_{0}^{2}$ and introduce a dimensionless parameter for
the amplitude of the Bell turbulence $\BellAmp = \sqrt{\bSqMean}/B_0$.
%
The turbulence correlation time $\taucor$, is estimated as the
short-scale mode vortex turnover time, while the amplitude of
the turbulent velocity is $\sqrt{\langle\mathbf{v}^{2}\rangle}$.
%
Then, the turbulent mixing length is defined as $\taucor
\sqrt{\langle\mathbf{v}^{2}\rangle}\approx\taucor
\sqrt{\xi\langle\mathbf{b}^{2}\rangle/(4\pi\rho)}= 2 \pi \xi/k_0$
where $\xi$ is a dimensionless parameter characterizing the
turbulence mixing length. In our numerical estimations below, we
assume $\xi\sim 5$ for the short-scale turbulence produced by Bell's
instability.

According to \citet{bell04}, the maximum growth rate of the
instability is at the wavelength $\lambda=4 \pi/k_0$. However, due
to the nonlinear dynamics of the Bell instability, the maximum of
the turbulent energy density may be at somewhat larger wavelengths
if the initial spectrum of the fluctuations subject to growth is a
declining function of the wavenumber $k$ \citep[c.f.,][]{zpv08,
rs09}.
We specify the scalings of the turbulent kinetic coefficients
with the dimensionless parameter $\xi$ as $k_0
\alpt/(4 \pi \rho) \approx 2 \pi \sqrt{\xi} \BellAmp v_a$,
%
$\kappa_t/B_0 \approx \pi \BellAmp $, and
assume that the minimum momentum of accelerated protons is
$p_{0}=mc$.
With these values, the
solution to the dispersion equation Eq.~(\ref{dispersMedium}) is
\begin{equation}\label{SolveDispersMedium}
\omega=\frac{1}{2}\left(-d\pm\sqrt{d^{2}-4c}\right),
\end{equation}
with
\begin{equation}\label{dCoeff}
d \equiv \mp
ikk_{0}\frac{\alpha_{t}}{4\pi\rho}\left[\frac{1}{2}A(x_{0})+\frac{3}{2}\right],
\end{equation}
and
\begin{equation}\label{cCoeff}
c \equiv -k^{2}v_{a}^{2}\pm kk_{0}v_{a}^{2}\left(1+\frac{
\kappa_t}{B_{0}}\right)\left[A(x_{0})-1\right].
\end{equation}

Below, we present simple estimations for frequencies, growing-rates
and polarizations for the fastest growing modes. As has long been
known, the standard resonant instability
\citep[e.g.,][]{be87,kulsr05} operates in the intermediate
magnetized regime $\eta^{-1} < kr_{g0} < 1$ in the lack of strong,
short-scale turbulence.
This resonant effect in the dispersion equation~(\ref{dispersMedium}) is
dominated by the imaginary part of the current response function
$A(x_{0})$ defined by Eqs.~(\ref{coeffANu},~\ref{sigmaCr}). The response
function was calculated for a power-law momentum distribution of CRs of
index $\alpha$ as defined in Eq.~\ref{disrNp}.  Modes with
different circular polarizations that are distinguished by the sign
$\mp$ in Eqs.~(\ref{coeffANu},~\ref{sigmaCr}) have the same growth rate
and with no short-scale turbulence (i.e., $\alpha_{t}=0$ and $\kappa_t
=0$), the frequencies of the two circularly polarized modes are
determined by
\begin{equation}\label{omegResonAlf}
\frac{\omega r_{g0}}{v_{a}}\approx \pm\left(1+i\right)
\sqrt{\frac{3\pi}{8}\left(\frac{1}{\alpha-2}-\frac{1}{\alpha}\right)k_{0}r_{g0}}
(kr_{g0})^{(\alpha-2)/2}
\ .
\end{equation}
This growth rate is plotted as the dashed curve in Fig.~\ref{parallel}
  for $k_0 \rgz=100$ and $\alpha=4$.

Now let's consider the effect of strong, short-scale Bell turbulence
on the resonant instability when $\alpha_{t}$ and $\kappa_t$ are
nonzero. The main contribution to the dispersion
equation~(\ref{SolveDispersMedium}) at $\xi\sim 5$ is due to the
coefficient $d$ in Eq.~(\ref{dCoeff}). Since the response function
$A(x_{0})\approx 1$ for $kr_{g0} < 1$, the growth rate can be
approximated as
\begin{equation}\label{OmResonAlf1}
\omega=i4\pi\sqrt{\xi}\BellAmp v_{a}k \ .
\end{equation}
For $kr_{g0} < 1$, only the mode with polarization $\delta
\mathbf{b}=\delta b(\mathbf{e}_{x}+ i\mathbf{e}_{y})$ is growing
while for the case $k \rgz > 1$, Bell's instability amplifies the
other mode with polarization $\delta \mathbf{b}=\delta
b(\mathbf{e}_{x}- i\mathbf{e}_{y})$.
The ratio of the kinetic energy density to the magnetic energy density
in the growing mode can be estimated from
\begin{equation}\label{estimEnerg1}
|\delta\overline{\mathbf{V}}(\mathbf{k})|^{2}\sim
\frac{1}{4\pi\rho_{0}}\left(\frac{3k_{0}r_{g0}(1+\pi
N_{B})}{64\pi\sqrt{\xi} N_{B}}\right)^{2}|\delta
\mathbf{b}(\mathbf{k})|^{2} \ .
\end{equation}

It should be noted that the helicity of the unstable, long-wavelength
mode studied above is opposite that of the short-scale Bell mode. This
provides, in principle at least, the possibility of balancing the global
helicity of the system by combining short and long-wavelength
modes. Care must be taken however, since recent numerical simulations
show a high saturation amplitude of the Bell mode making a nonlinear
analysis necessary to address the helicity balance issue. The
estimations given above are valid in the intermediate regime and provide
simple analytical approximations to the growth rates shown in
Fig.~\ref{parallel} for $k \rgz \lsim 1$. To turn to the \hydro\ regime
$k \rgz < \eta^{-1}$ (considered in Section \ref{Hydro}), one should
just change $(1 + \kappa_t/B_0)$ to $\kappa_t/B_0$
in Eq.~(\ref{cCoeff}), so that
\begin{equation}\label{cCoeffNu}
c=-k^{2}v_{a}^{2}\,\pm\, kk_{0}v_{a}^{2} \kappa_t
\left[A(x_{0})-1\right]/B_0
\ ,
\end{equation}
and then substitute this in Eq.~(\ref{SolveDispersMedium}).

The coefficient given by Eq.~(\ref{cCoeffNu}) (or more exactly the
imaginary part of Eq.~\ref{coeffANu1}) dominates the dispersion equation
Eq.~(\ref{SolveDispersMedium}) if $x_{0}\ll 1$ and $\eta$ is finite.
Because of the $\mp$ sign in the imaginary part of
Eq.~(\ref{coeffANu1}), both circular polarizations will be growing with
the same growth rate given by
\begin{equation}\label{gam1}
\gamma\approx
\sqrt{\frac{\pi\BellAmp}{2}}\sqrt{\frac{kk_{0}}{\eta}}v_{a}.
\end{equation}
The transition from the regime described by Eq.~(\ref{OmResonAlf1}) to that
described by Eq.~(\ref{gam1}) takes place at
\begin{equation}\label{x0}
x_{0}\sim\frac{1}{\eta}\frac{k_{0}r_{g0}}{32\pi\xi\sqrt{\BellAmp}} \
.
\end{equation}

\subsection{Long-wavelength current driven
modes in the hydrodynamical regime ($kr_{g0} < \eta^{-1}$)}\label{Hydro}

The ponderomotive force
$\langle (\jCRVec -e \nCR \mathbf{v}) \times \mathbf{b}\rangle/(c \rho)$
in the mean-field momentum equation of the background plasma
Eq.~(\ref{largeEqMotion0}), is due to the momentum exchange between the
background plasma and cosmic rays. Contrary to the short-wavelength
regime, the cosmic-ray current response on the magnetic fluctuations is
essential in the long-wavelength regime, $kr_{g0} < 1$, and results in a
non-negligible ponderomotive force.
If the perturbation wavelength is longer than the CR mean free path,
$\Lambda$, or in other words if $kr_{g0} < \eta^{-1}$, then the
hydrodynamic approximation can also be applied to the cosmic-ray
dynamics. Then, the momentum density $\PrVec$ and the momentum density
flux tensor $\Ptensor$ of the CR-fluid, defined as
\begin{equation}\label{R1}
\Pi^{(r)}_{\alpha\beta}=\int v_\alpha p_\beta \fCR d^{\,3}p
      =\rho_r u_\alpha^{(r)}u_\beta^{(r)}
      + \PCR \delta_{\alpha\beta},
\end{equation}
can be approximately derived in closed form. Here $\fCR$ is the CR
distribution function. Then, the CR momentum equation, derived as
a moment of the kinetic equation, takes the form
\begin{equation}\label{hMomCR1}
\frac{\partial\mathbf{P}_{\alpha}^{(r)}}{\partial
 t}+\nabla_{\beta}\cdot{\Pi}_{\alpha\beta}^{(r)}
 = \frac{1}{c}[(\JCR \times\mathbf{B}) + e \nCR
 \mathbf{E}]_{\alpha}.
\end{equation}
Equation~(\ref{hMomCR1}), averaged over the fluctuations with
scales below the CR mean free path and taking into account
Eq.~(\ref{Jcrtotal}), has the form
\begin{eqnarray}\label{hMomCR2}
& &
\left\langle\frac{\partial\mathbf{P}_{\alpha}^{(r)}}{\partial
 t}+\nabla_{\beta}\cdot{\Pi}_{\alpha\beta}^{(r)}\right\rangle
 =
 \nonumber \\
& &
=\frac{1}{c}[(\jCRmeanVec - e \nCR \Vbar) \times \Bbar +
 \left\langle(\jCRVec -e \nCR \mathbf{v})\times\mathbf{b}\right\rangle]_{\alpha}.
\end{eqnarray}

 The CR distribution in Eq.~(\ref{R1}) is nearly isotropic for scales
 larger than the CR mean free path $kr_{g0} < \eta^{-1}$. Then, the
 isotropic cosmic-ray pressure dominates in $\Pi^{(r)}_{\alpha\beta}$
 and using Eq.~(\ref{hMomCR2}) one may write a simplified
 Eq.~(\ref{largeEqMotion0}) in the form
\begin{eqnarray}\label{largeEqMotion1}
& &
\frac{\partial\delta\mathbf{\overline{V}}}{\partial t}
=-\frac{1}{\rho}\nabla \left(\delta P+\delta P_{cr}\right)
+\frac{1}{4\pi\rho}((\nabla\times\delta\mathbf{b})\times
B_{0}\mathbf{e}_{z})+
\nonumber \\
& &  +\frac{\kappa_t}{c\rho}\left(\left[\delta
j^{cr}_{x}-g^{\,\prime}\delta
b_{x}\right]\mathbf{e}_{y}-\left[\delta
j^{cr}_{y}-g^{\,\prime}\delta b_{y}\right]\mathbf{e}_{x}\right) \ .
\end{eqnarray}
In the rest frame of the unperturbed shock precursor, the velocity
has only the perturbed component
$\mathbf{\overline{V}}=\delta\mathbf{\overline{V}}$.
The term $\sim e \nCR \delta\mathbf{\overline{V}}$ is small compared
with $g^{\,\prime}\delta \mathbf{b}$ if $v_{\rm{ph}}/v_s \ll 1$,
and it is omitted in the dispersion relation.

To study the dispersion properties of the CR current driven
modes in the long-wavelength regime, we use the mean-field dynamic
equations averaged over short-scale fluctuations, Eqs.~(\ref{MediumInd1}),
and (\ref{largeEqMotion1}), and the mass continuity equation:
\begin{equation}\label{eqCont1}
    \frac{\partial\rho}{\partial t}+\nabla(\rho \Vmean)=0
\ .
\end{equation}
The equation of state for the background plasma is assumed to be
adiabatic, i.e.,
\begin{equation}\label{eqP1}
    \frac{\partial P}{\partial t}+\left( \Vmean \nabla
    \right)P +\gamma P\nabla \Vmean = 0,
\end{equation}
where $\gamma$ is the adiabatic index.
Using
Eq.~(\ref{eqCont1}) and Eq.~(\ref{eqP1}), one obtains
\begin{equation}\label{gradpressBG}
\nabla\delta P=a_0^{2}\nabla \delta\rho,
\end{equation}
where $a_0 = \sqrt{\gamma P_0/\rho_0}$
is the sound speed of the background plasma.

To get the dispersion equation one must substitute the expressions
$\delta\mathbf{b}=\mathbf{\overline{B}}-\mathbf{B}_{0}$,
$\delta\Vmean = \Vmean $, $\delta\mathbf{j} = \jCRmeanVec  - \jpar$
into Eqs.~(\ref{MediumInd1}) and  (\ref{eqCont1}). Then, using
Eqs.~(\ref{curCRxy}) and  (\ref{gradpressBG}), by  neglecting the
terms that are small at $x < x_0$ (where $x_0$ is defined by
Eq.~\ref{x0}), we finally get the linear dispersion relation for the
perturbations with wavelengths larger than the mean free path of the
cosmic-ray particle $\Lambda$, i.e.,
%
\begin{eqnarray}\label{Displarge0}
  & &
(\omega^{2}-k_{\parallel}^{2}{v}_{a}^{2}-ik_{\parallel}B_{0}W_{0})\{\omega^{4}-(a_{0}^{2}+v_{a}^{2})k^{2}\omega^{2}+
 \nonumber \\
    & &
+k_{\parallel}^{2}k^{2}v_{a}^{2}a_{0}^{2}
-ik_{\parallel}B_{0}[(\omega^{2}-a_{0}^{2}k_{\parallel}^{2})W_{3}
+k_{\perp}k_{\parallel}a_{0}^{2}W_{4}]\}-
 \nonumber \\
    & &
 -k_{\parallel}^{2}B_{0}^{2}W_{1}[(\omega^{2}-k_{\parallel}^{2}a_{0}^{2})W_{2}-k_{\perp}k_{\parallel}a_{0}^{2}W_{5}]=0,
\end{eqnarray}
where we assumed $k_{x}=0$, $k_{y}=k_{\perp}$, and
\begin{eqnarray}\label{W0}
& &
W_{0}=g^{\,\prime}\frac{
\kappa_t}{c\rho_{0}}\frac{1}{\eta}D_1(k_{\parallel},k_{\perp}),
\nonumber\\
& &
W_{1}=g^{\,\prime}\frac{
\kappa_t}{c\rho_{0}}\left[D_1(k_{\parallel},k_{\perp})-1\right],
\nonumber\\
& &
W_{2}=g^{\,\prime}\frac{
\kappa_t}{c\rho_{0}}\left[D_1(k_{\parallel},k_{\perp})-1\right]
-g^{\,\prime}\frac{B_{0}}{c\rho_{0}}\frac{1}{\eta^{2}}D_2(k_{\parallel},k_{\perp}),
\nonumber\\
& &
W_{3}=g^{\,\prime}\frac{\kappa_t}{c\rho_{0}}\frac{1}{\eta}[D_1(k_{\parallel},k_{\perp})+D_2(k_{\parallel},k_{\perp})]+
\nonumber\\
& &
+g^{\,\prime}\frac{B_{0}}{c\rho_{0}}\frac{1}{\eta}D_2(k_{\parallel},k_{\perp}),
\nonumber\\
& &
W_{4}=g^{\,\prime}\frac{B_{0}}{c\rho_{0}}D_3(k_{\parallel},k_{\perp}),
\nonumber\\
& &
W_{5}=g^{\,\prime}\frac{B_{0}}{c\rho_{0}}\frac{1}{\eta^{2}}D_3(k_{\parallel},k_{\perp}).
\end{eqnarray}
The angular dependence of the coefficients $W_{i}$ for $i$= 0, \dots,
4 is determined by
\begin{eqnarray}\label{D1}
& &
 D_1(k_{\parallel},k_{\perp}) =
\frac{k_{\parallel}^{2}}{k_{\parallel}^{2}+\frac{k^{2}}{\eta^{2}}},\,\,
D_2(k_{\parallel},k_{\perp}) =
\frac{k_{\perp}^{2}}{k_{\parallel}^{2}+\frac{k^{2}}{\eta^{2}}},\,\,
\nonumber\\
& &
 D_3(k_{\parallel},k_{\perp})
=\frac{k_{\perp}k_{\parallel}}{k_{\parallel}^{2}+\frac{k^{2}}{\eta^{2}}}
\ .
\end{eqnarray}
All of the coefficients $W_{i}$ given above are proportional to
$g^{\,\prime}$ and thus they vanish in the absence of the
unperturbed cosmic-ray current, i.e., if $\jzCR=0$. The
corresponding terms are therefore responsible for the current driven
modes. The coefficients have different asymptotic behaviors at $\eta
\gg 1$. The coefficients $W_{0}$,  $W_{3}$, and  $W_{4}$ scale
$\propto \eta^{-1}$, while all of the others are $\propto
\eta^{-2}$. It is important to note, however, that  contrary to
$W_{3}$, $W_{4} \propto k_{\parallel}$ and $W_{0}\propto
k_{\parallel}^2$.

\begin{figure}
\centering {
 \rotatebox{0}{
{\includegraphics[height=8.5cm, width=12.5cm]{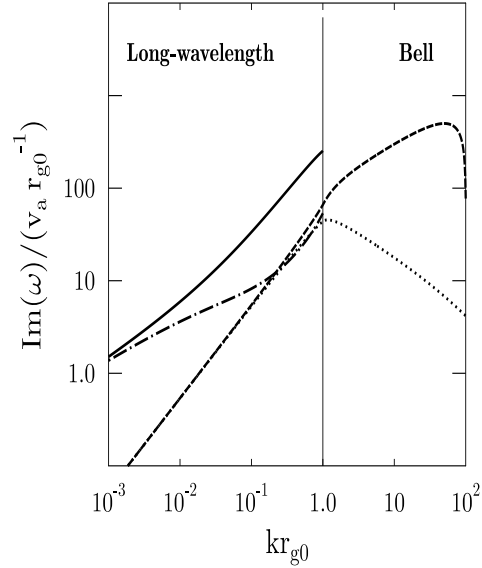} }}}
\caption{The figure shows growth rates of the parallel propagating modes as a
    function of the wavenumber to illustrate the effect of short-scale
    turbulence on the long-wavelength instability.
    Equation~(\ref{SolveDispersMedium}) was solved numerically to generate
  the curves. The model parameters
    are $k_{0}r_{g0} = 100$ and $\alpha = 4.0$. The solid and dot-dashed
    curves are simulated for two modes in the model with short-scale
    turbulence of $\xi =5$ and $\eta = 10$ to demonstrate the behavior
    of the modes in the intermediate regime. For comparison, the dashed
    and dotted curves are calculated for the model without the
    short-scale turbulence, i.e., with $\BellAmp =0$ and $\eta
    \rightarrow \infty$ \citep[c.f.,][]{bell04}.}\label{parallel}
\end{figure}

The growth rates depend on the mode propagation angle and we first
consider parallel propagating modes. As discussed above, for modes
propagating parallel to the initial magnetic field, the growth rate is
\begin{equation}\label{incr2}
\gamma_0(k) \approx
\sqrt{\frac{\pi\BellAmp}{2}}\sqrt{\frac{kk_{0}}{\eta}}v_{a} \ ,
\end{equation}
and these modes have the fastest growth rates for the Bohm diffusion
regime with $\eta \sim 1$.
In Fig.~\ref{parallel} we illustrate the effect of short-scale
turbulence on the long-wavelength instability for a particular set
of parameters. The dashed and dotted curves show the result
without short-scale turbulence. The two curves show two growing
short-scale modes, two other modes in Bell's dispersion equation
are not growing and aren't shown.
The departure of the solid and dot-dashed curves from the linear
dependence (evident flattening) at small $k$ is due to the transition to
the \hydro\ regime when $kr_{g0} < \eta^{-1}$. It is clearly seen in
Fig.~\ref{parallel} that the resonant growth rates calculated for the
model with short-scale turbulence are well above the standard resonant
growth rates shown by the dashed line at $k r_{g0}<$ 1.

\begin{figure}
\centering {
 \rotatebox{0}{
{\includegraphics[height=8.5cm, width=9.5cm]{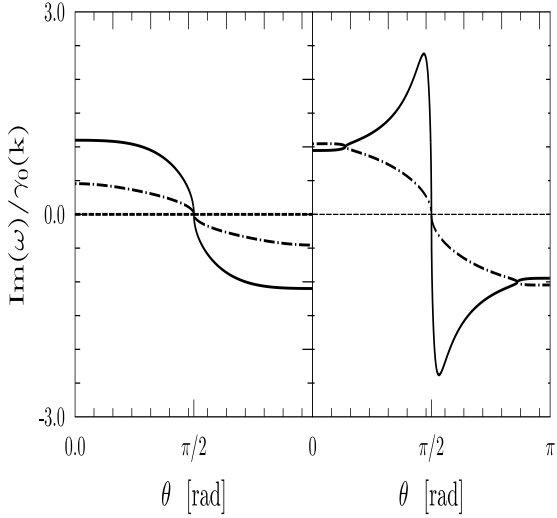} }}}
\caption{The two panels show the angular dependence of the growth
rates of the long-wavelength unstable modes in the \hydro\ regime
where $kr_{g0} < \eta^{-1}$. In the left panel $\eta = 1$, while
$\eta = 10$ in the right panel. The model parameters are
$k_{0}r_{g0} = 100$, $\alpha = 4.0$, and the plasma parameter $\beta
= a_0^2/v_a^2 = 1.0$.  The solid and dot-dashed curves show the two
unstable modes.
The normalizing parameter $\gamma_0$ is determined by
Eq.~(\ref{incr2}).}\label{angular}
\end{figure}

If $\eta >1$, scattering is less efficient than the Bohm limit and
the maximum growth rate occurs for obliquely propagating modes. The
non-parallel propagating modes are driven by both the cosmic-ray
current and the cosmic-ray pressure gradient (i.e., the diffusive
part of the cosmic ray current).  An analysis of the relative
contributions of the corresponding terms in $W_3$ defined in
Eq.~(\ref{W0}) using Eq.~(\ref{curCRxy}), shows that, at the maximum
growth rate, the relative contribution of the cosmic-ray pressure
gradient to that of the cosmic-ray current in the shock precursor is
$\propto \pi\BellAmp$. Therefore, the cosmic-ray current
contribution is larger if the short-scale Bell turbulence is
strong.
In Fig.~\ref{angular}, the angular dependence of the \hydro,
long-wavelength modes is illustrated for a finite temperature plasma
with the parameter $\beta = a_0^2/v_a^2 = 1.0$. One can see that, in
contrast to the short-scale Bell instability, the modes propagating
along the unperturbed magnetic field are growing while the modes
propagating in the opposite direction are damped. It is also clear
in Fig.~\ref{angular} that the fastest growing modes are propagating
nearly perpendicular to the unperturbed field $\mathbf{B_0}$ for
$\eta = 10$, a typical result for $\eta \gg 1$.

To get the propagation angle $\Tmax$ of the mode of maximum growth
for  $\eta>1$, one needs to find the  maximum of the expression
$\kpar W_3 (\kpar/k = \cos{\theta})$, i.e.,
\begin{equation}\label{thetaMax}
\cos{\Tmax} = 1 / \eta ,
\end{equation}
and the maximum growth rate at $\eta\gg 1$ is determined by
\begin{equation}\label{incr8}
\gamma(k)\approx\sqrt{\frac{\pi\BellAmp}{4}}\sqrt{kk_{0}}v_{a}.
\end{equation}

The results were obtained assuming $v_{\rm{ph}}/v_s <<1$. In our
case $\gamma(k)\sim\omega(k)$, and therefore from Eq.~\ref{incr8}
one may get the validity condition in the form
$\displaystyle\sqrt{\frac{\pi\BellAmp}{4}}\sqrt{\frac{k_{0}}{k}}\,M_{\rm{a}}^{-1}<<1$
(where $M_{\rm{a}}$ is the \alf Mach number of the shock).

 Consider the polarization of a mode propagating along the
direction giving the maximum growth rate for $\eta\gg 1$ (i.e.,
$k_{\perp}\gg k_{\parallel}$, see Fig.~\ref{angular}). The amplitude
of the magnetic field along the initial magnetic field, $\delta
b_{z}$, in this mode, exceeds the transverse magnetic field
perturbations $\delta b_{x,y}$. The maximum velocity component
$\delta\overline{V}_{y}$ is in the plane determined by the wave
vector and the initial magnetic field
\begin{equation}\label{IndDelta}
\delta\overline{V}_{y}=\frac{\omega}{v_{a}k_{\perp}}\frac{\delta
b_{z}}{\sqrt{4\pi\rho_{0}}}\approx
-\frac{\omega}{v_{a}k_{\parallel}}\frac{\delta
b_{y}}{\sqrt{4\pi\rho_{0}}},
\end{equation}
and the energy density of the mode is dominated by the kinetic energy
\begin{equation}\label{estimEnerg}
|\delta\overline{V}_{y}(\mathbf{k})|^{2}\sim
\frac{1}{4\pi\rho_{0}}\frac{\pi N_{B}}{4}\frac{k_{0}}{k}|\delta
b_{z}(\mathbf{k})|^{2},
\end{equation}
since $k_{0}\gg k$, and we assumed $|k_{\perp}|\approx k$.

The modes are compressive and from the continuity equation one
can write
\begin{equation}\label{Nepr}
\frac{\delta\rho}{\rho_{0}}=
\frac{k_{\perp}\delta\overline{V}_{y}+
k_{\parallel}\delta\overline{V}_{z}}{\omega}
\approx\frac{k_{\perp}\delta\overline{V}_{y}}{\omega} .
\end{equation}
Then, using the induction equation given above, the equation $\nabla
\delta\mathbf{b}=0$, $k_{\perp}> k_{\parallel}$, and
$\delta\overline{V}_{y}\gg \delta\overline{V}_{z}$, one can estimate
 $\displaystyle
\left|\frac{\delta\rho}{\rho_{0}}\right|\sim
\left|\frac{\delta\mathbf{b}}{B_{0}}\right|$. The complex frequency
$\omega$ is dominated by the growth rate $\gamma(\mathbf{k})$ in the
regime under consideration. The analysis above was performed using
the turbulent kinetic coefficients determined by Bell's turbulence
that are valid if
\begin{equation}\label{thermEf}
\left(\frac{v_{a}}{v_{Ti}}\right)^{2} > k_{0}r_{g0}\frac{v_{a}}{c},
\end{equation}
where $v_{Ti}$ - is the thermal ion velocity.
A thorough discussion of the
effects of a hot plasma on the short-scale modes was done by
\citet{ze10}.

We emphasize that the growth rates obtained here account for both the CR
current and the CR pressure gradient in the presence of short-scale Bell
turbulence.  In the presence of the short-scale fluctuations, the
momentum exchange between the CRs and the flow in the \hydro\ regime,
$kr_{g0} < \eta^{-1}$, results in a ponderomotive force proportional to
the CR current (the last terms in the mean-field momentum
equation~\ref{largeEqMotion1}). As a result, there exist transverse
growing modes with wavevectors along the initial magnetic field with
growth rates that are proportional to the turbulent coefficient
$\kappa_{t}$ defined in Appendix \ref{EMotion}.
In the opposite regime with $\eta \gg 1$, the fastest growth rates are
for the modes having wavevectors nearly transverse to the
initial magnetic field. These modes are compressible. If the short-scale
Bell turbulence is absent (i.e.,  when the turbulent coefficients
vanish), unstable acoustic modes are produced, as has been studied earlier by
\citet{chalov88}.

\section{Discussion and Conclusions}\label{Disc}
Collisionless shocks are complex phenomena where a number of relaxation
processes are involved that redistribute the bulk ram kinetic
energy into individual superthermal particles and magnetic fields. Strong
astrophysical shocks can transfer a sizable fraction of the ram energy
from thermal particles to extremely relativistic ones and the great
challenge for modeling these shocks comes from this wide dynamic range.
Furthermore, if the particles have a wide range in momentum,
the self-generated magnetic turbulence must have a correspondingly wide
range in wavelengths.  Thus far, no exact treatment of this process with
plasma simulations or other methods has been
possible and approximate techniques must be used.

In this paper, we derive a mechanism for long-wavelength magnetic
fluctuation growth in the presence of a cosmic-ray current and the
short-scale magnetic turbulence produced by Bell's instability.
We use the anisotropic cosmic-ray
distribution, interacting with strong Bell-like turbulence with scales
below the CR gyroradii, to calculate the growth rate of fluctuations
with wavelengths longer than the CR gyroradius or mean free path.
As we have emphasized, the power in the longest wavelength
turbulence is critical for determining the highest energy CRs a given
shock can produce.

The algebra needed to describe the mechanism is long but, schematically,
this is what happens: The $\JCR \times \mathbf{B}/c$ force from the CR
current drives the Bell short-scale instability at scales below the CR
gyroradius. Strong, short-scale turbulence is produced if the $\JCR
\times \mathbf{B}/c$ force is large enough to dominate the magnetic
field tension in the momentum equation (\ref{eqMotiontot1}).
The short-scale turbulence influences the large-scale dynamics through
the ponderomotive forces imposed on the plasma by the turbulence and the
CR current.
To derive the mean ponderomotive force one must average
the momentum equation over the ensemble of short-scale fluctuations.
When this is done, it is seen that there is a contribution to the
ponderomotive force (Eqs.~\ref{largeEqMotion0}, \ref{preobr1},
\ref{preobr2}). In the intermediate regime, to first order in the
amplitude of the long-wavelength fluctuations, this contribution is
$
(1 + \kappa_t/B_0) \jCRmean \times \Bbar/c$,
where $\mathbf{\overline{B}}=B_{0}\mathbf{e}_{z}+\delta\mathbf{b}$
and $\jCRmean = \jzCR \mathbf{e}_{z} + \delta\jCR$.
The effect of the short-scale turbulence appears here through the
turbulent coefficient $\kappa_{t}/B_0$.
We find that the turbulent ponderomotive force is large enough in
both the intermediate and hydrodynamical regimes, and that the CR
current response in the long-wavelength regime can no longer be
neglected (see Section~\ref{CRCR}). The current cannot be treated as
a fixed external parameter, as is normally done for the short-scale
Bell instability.

The CR current response (Eq.~\ref{curCRxy}) on the magnetic
turbulence is derived here using a kinetic equation with a
simplified collision operator.  With this approximation,  the
ponderomotive force  results in a long-wavelength instability in a
way similar to Bell's instability. The angular dependence of the
growth rate in the hydrodynamical regime depends on the
dimensionless collision strength $\eta$, as is shown in
Fig.~\ref{angular}.

The method we use to average the ponderomotive and
  electromotive forces in the presence of the cosmic ray current
  is a generalization of the mean field method used
  in dynamo theory \citep[see][ for a recent review]{brandenburg09}.  By
  introducing a parameterized relaxation time $\taucor$ for the triple
  correlations of the short-scale field fluctuations, we are able to
  obtain the mean field equations in closed form.
%
The turbulent transport coefficients (Eqs.~\ref{turbcoef1} and
\ref{turbcoef2}) needed to determine the pondermotive force in the
mean field momentum equation are written in terms of a dimensionless
turbulent mixing length that is dependent on $\taucor$ and the
dimensionless turbulence amplitude $N_B$.

In the intermediate regime ($\eta^{-1} <
kr_{g0} < 1$), the fastest growing modes (i.e.,
Eq.~\ref{OmResonAlf1}) appear due to the CR current contribution to the
mean electromotive force in the averaged magnetic induction equation
(\ref{MediumInd2}). The CR contributions are proportional to
the turbulent transport coefficient $\alpha_{t}$.  The amplification of
long-wavelength magnetic fields in this  regime is reminiscent of the
large-scale magnetic field dynamo model that is widely discussed in the
literature \citep[see e.g.][ for a review]{bs05}.

The fastest growing long-wavelength mode in the intermediate regime
has non-zero helicity with a sign opposite to the short-wavelength
Bell modes.
In the hydrodynamical regime ($k \rgz < \eta^{-1}$), the two modes
with similar growth rates (the solid and dot-dashed curves in
Fig.~\ref{parallel}) dominating the regime $k \rgz < \eta^{-1}$,
also have opposite helicity.
While is is not possible to draw exact conclusions
for the properties of the strong turbulence with our linear analysis,
the growth rates of the long-wavelength turbulence we obtain are
comparable to those of Bell's instability for a reasonable range of
parameters (compare the maxima of the solid and dashed curves in
Fig.~\ref{parallel}).  This suggests that it may by possible to overcome
a fundamental problem and balance the overall helicity with these two
instabilities, but non-linear analysis is needed to address the issue.

The nonlinear, long-wavelength instability requires an accounting of
the CR response to the short-scale turbulence.  Here, we have
exploited the fact that the process of long-scale turbulence growth
in the presence of short-scale motions and magnetic field
fluctuations, has some analogy with the dynamo theory reviewed
recently by \citet{brandenburg09}. The key ingredient in our model,
however, is the anisotropic CR distribution.

Previous nonlinear work on the short-scale Bell instability using MHD
simulations \citep[e.g.,][]{bell04,zp08,zpv08} assumed a fixed CR
current as an external parameter. These models studied the spectral
evolution in the short-scale range as well as the transformation of
the turbulence through the subshock.  The evolution of the Bell
modes downstream from the shock was addressed by
\citet{pelletierea06,marcowithea06}.

 In DSA, the shock precursor containing the CR
current has a scale length $l_f(p) \approx c \Lambda(p)/v_s$.
Therefore, the growing modes are advected through the instability
region on a time scale
\begin{equation}\label{timeDrift}
\tau_d(p) \approx c\Lambda(p)/(3v_{s}^{2})
\ .
\end{equation}
 To amplify a mode of wavenumber $k$ by a  factor of a few, one
 needs the growth rate to satisfy
\begin{equation}\label{timeDrift1}
\gamma(k)\cdot \tau_{d}> 1.
\end{equation}

The growth rates  can be estimated for a particle diffusion model
widely used in DSA \citep[e.g.,][]{be87}, where the particle mean free
path is
\begin{equation}\label{probeg}
\Lambda \left( {p} \right) = \eta \,r_{g} \left( {p} \right) = 3
\cdot 10^{12}\eta \,\,\left( {{\frac{{{\rm B}_{0}}} {{{ 1 \mu \mathrm{G}}{\rm
ñ}}}}} \right)^{ - 1}\left( {{\frac{p}{{m_{p} c}}}} \right) \mathrm{cm} .
\end{equation}
Consider the energy density, $\ECR$, for a power-law
CR distribution
as expected in test-particle DSA. This is the same as
given by Eq.~(\ref{disrNp}) only here we normalize the CR number density
to $\nCR$.
Then, if the power-law index is $\alpha$, the CR energy density
can be written as
%
\begin{equation}\label{Ecr1}
\ECR =  cp_{0} \nCR \Phi(\alpha),
\end{equation}
where
\begin{equation}\label{paraPhi}
\Phi(\alpha) =
\left\{%
\begin{array}{ll}
    \displaystyle
    \frac{\alpha-3}{\alpha-4}
\left[ 1-\left(\frac{p_{m}}{p_{0}}\right)^{4-\alpha}\right],\,
\mathrm{for}  \ \alpha>4 \\
    \displaystyle \ln \left( {{\frac{{p_{m}}} {p_{0}}}} \right), \,
\mathrm{for } \ \alpha=4 \\
    \displaystyle
\frac{\alpha-3}{4-\alpha}
\left[ \left(\frac{p_{m}}{p_{0}}\right)^{4-\alpha}-1\right] ,
    \,
\mathrm{for}  \ \alpha<4,
\end{array}%
\right .
\end{equation}
and $p_m$ is the maximum CR momentum and $p_0$ is the minimum.
It is convenient to write
\begin{equation}\label{Ecr2}
\ECR = \epCR {\frac{{n_{p} m_{p} v_{s}^{2}}} {{2}}},
\end{equation}
where $\epCR$ is the fraction of the far upstream shock ram pressure
transferred to the cosmic-ray energy density. Note that $\epCR$ varies
with precursor position and with the minimum CR momentum, $p_0$.
Then the maximum growth rate achievable in the model, for a mode of
wavenumber $k= 2\pi/\Lambda$ propagating at the angle $\cos\theta
\approx \eta^{-1}$ to the initial magnetic field, is given by
\begin{eqnarray}\label{incrTimeDriftMax}
& & \GamMax \tau _{d} \approx 1.7 \left(\frac{\eta
\BellAmp}{10}\right)^{1/2} \left(\frac{\Phi}{10}\right)^{-1/2}\times
   \nonumber \\
  & &
\times\left(\frac{\epCR}{0.1}\right)^{1/2}
\left(\frac{v_{s}}{0.01c}\right)^{-1/2} .
\end{eqnarray}
Equation~(\ref{incrTimeDriftMax}) demonstrates that, for parameters
typical of young SNRs, and for acceleration efficiencies that are
high enough (i.e., $\epCR \gsim 0.1$), long-wavelength fluctuations
can be strongly amplified. The validity requirement given by
Eq.~(\ref{thermEf}) is fulfilled for most of the conditions expected
in young galactic SNRs.
%

In DSA,  particle
spectra far upstream in the shock precursor close to the escape boundary
can be approximated by
Eq.~(\ref{disrNp}) where the minimal momentum $p_{0}$ can be just a
few times smaller than $p_m$.
In contrast, just upstream from the subshock, the
minimal momentum is
much lower, i.e.,  $p_{0} \sim m_pc$.
Therefore, for $p_m \gsim 10^5 m_pc$, the particle mean free path
$\Lambda \lsim 10^{18}$ cm and fluctuations with wavelengths larger than
$\Lambda$ would grow with a time scale on the order of 1,000 yr for a
shock of velocity $v_{s}\sim 0.01c$.

Besides increasing the maximum CR proton energy a shock can produce, the
presence of strong, long-wavelength magnetic field fluctuations can
affect synchrotron emission from relativistic electrons \citep[see,
e.g.,][]{vl03,bambaea05,uchiyamaea07,vink08}.
Temporal, spatial, spectral, and polarization features seen in \syn\
emission from the shells of young supernova remnants will all be
modified to some extent by long-wavelength turbulence. The
observation of these features can provide unique information on the
properties of the long-wavelength fluctuations \citep[see,
e.g.,][]{bue08,bubhk09}.
In case of $\eta >1$, and for the quasi-parallel shocks considered here,
  the growing long-wavelength magnetic field fluctuations are
  propagating obliquely to the initial magnetic field. This means they
  must be accompanied by plasma density fluctuations that are
  interacting with the shock front, an effect that potentially can be
  studied through the optical line observations of SNRs \citep[see
  e.g.,][]{raymondea10}.

The interpretation of optical observations might also be
  influenced, at least for the  quasi-parallel shocks we studied here.
 For $\eta >1$,  the growing,
long-wavelength magnetic field fluctuations are propagating
obliquely to the initial magnetic field and they must be accompanied by
plasma density fluctuations. These density fluctuations might produce
  observable features in optical line observations of SNRs \citep[see,
e.g.,][]{raymondea10}.
%







\bibliographystyle{mn2e}
\bibliography{mf_}

\appendix

\section{Linear response of the cosmic ray current}\label{CRCR}

A small perturbation $\delta \mathbf{b}$ imposed on the local mean
magnetic field in the form of a plane monochromatic wave results in a
linear response of the cosmic-ray current $\delta \jCR$.  General
expressions for the linear response of a background plasma at rest with
an arbitrary particle distribution, were presented by
\citet[][]{kt73}. Kinetic equations will be used here to derive the
linear response of the cosmic-ray current. For the unperturbed CR
distribution, we consider distributions that are typical for the
diffusive shock acceleration model \citep[e.g.,][]{be87}. In this case,
the momentum distribution of accelerated particles allows a local
approximation to a weakly inhomogeneous (on scales of the order of
the cosmic-ray gyroradii) distribution given by
\begin{equation}\label{distrFuncCr}
f_{0}(\mathbf{p})=
\frac{\nCR}{4\pi}N(p)\left[1+\frac{3v_{s}p_{z}}{cp}\right],
\end{equation}
where $v_{s}$ is the drift velocity of the accelerated particles
along the mean magnetic field $\mathbf{B}_{0}$  relative to the
background plasma ions, and  $N(p)$ is the isotropic part of the
cosmic-ray particle momentum distribution of the power-law form
\begin{equation}\label{disrNp}
N(p)=\frac{(\alpha-3)p_{0}^{\alpha-3}}{p^{\alpha}},~~~ p_{0}\leq
p\leq p_{m},~~~\alpha >3,
\end{equation}
normalized as $\displaystyle\int_{p_0}^{p_m}N(p)p^{2}dp=1$.

Actually, in DSA the particle distribution is inhomogeneous on the scale
$l_f \approx \Lambda c/v_s$ which is much larger than the particle
gyroradius in the \nonrel\ shocks considered here.
Thus, in our local analysis, the minimum momentum of the particle
distribution $p_0$ in Eq.~(\ref{disrNp}) is position dependent. In
the far upstream region, $p_0$ approaches $p_m$ making the
distribution given by Eq.~(\ref{disrNp}) rather narrow. In contrast,
at the shock front $p_0 \approx m_pc$ and the distribution is much
broader.
%

The particle distribution function of energetic particles,
$f(\mathbf{r},p,\theta,\varphi,t)$, disturbed by a magnetic field,
satisfies the kinetic equation
\begin{equation}\label{41}
\frac{\partial f}{\partial t}+\mathbf{v}\cdot\frac{\partial
f}{\partial\mathbf{r}}+e\mathbf{E}\cdot\frac{\partial
f}{\partial\mathbf{p}}-\frac{ec}{\mathcal{E}}(\mathbf{B}_0+\mathbf{b})
\cdot\widehat{\mathbf{\mathcal{O}}}f=I[f],
\end{equation}
where  $\mathbf{E}$ and $\mathbf{b}$
are the field amplitudes of the imposed MHD disturbance, and
$\widehat{\mathbf{\mathcal{O}}}$ is the momentum rotation operator,
defined by
\begin{equation}\label{a5}
\widehat{\mathbf{\mathcal{O}}}=\mathbf{p}\times\frac{\partial
}{\partial\mathbf{p}}
\ .
\end{equation}



\subsection{Perturbations propagating
along the initial magnetic field}\label{CRC11}
Let $\delta f$ be the perturbation of the distribution function due to
imposed harmonic perturbations of electromagnetic fields
$\mathbf{\delta b}$ and $\mathbf{E}\sim exp(i\mathbf{k}\cdot\mathbf{r}-i\omega t)$
propagating along the initial magnetic field direction
$\mathbf{k}=k\mathbf{e}_{z}=\mathbf{B}_{0}/B_{0}$.
We introduce the CR scattering rate by magnetic fluctuations, $\nu$,
assuming the simplest form for the CR collision operator in
Eq.~(\ref{41}), i.e.,
$I[f]=-\nu (\delta f- \Delfiso)$,
where $\Delfiso$ is the isotropic part of the disturbed
distribution function. Then, $\delta f$ satisfies
%
%
\begin{eqnarray}\label{kinEqFurCr}
& & -i(\omega-kv_{z})\delta
f+\frac{e}{c}\left(\mathbf{v}\times\mathbf{B}_{0}\right)\frac{\partial
\delta f}{\partial\mathbf{p}}+\nu\delta f=
\nonumber \\
  & &
=-e\left(\mathbf{E}+\frac{\mathbf{v}\times\delta\mathbf{b}}{c}\right)\frac{\partial
f_{0}}{\partial\mathbf{p}}.
\end{eqnarray}

In the case of parallel propagation, it is convenient to distinguish
between two circular polarization modes $\delta \mathbf{b}=\delta
b(\mathbf{e}_{x}\pm i\mathbf{e}_{y})$.
Then, the CR current response from Eq.~(\ref{kinEqFurCr}) takes the form
%
\begin{eqnarray}\label{CurrentCr1}
& &
 \delta \jCR \approx\mp\frac{3}{4}\mathbf{E}\frac{i}{\omega}e
\nCR \frac{c}{B_{0}} v_{s}k\times
\nonumber \\
  & &
\times\int_{0}^{\infty}\int_{-1}^{1}\frac{(1-t^2)dt}{1\mp xt\pm
ia}N(p)p^{2}dp,
\end{eqnarray}
%
where $x=\displaystyle\frac{kcp}{eB_{0}}$,
$a=\displaystyle\frac{\nu}{\Omega}=\frac{1}{\eta}$, and
$\Omega=\displaystyle\frac{eB_{0}}{p}$.
From Maxwell's equations one obtains
$\displaystyle\mathbf{E}=E(\mathbf{e}_{x}\pm
i\mathbf{e}_{y})=\pm i\frac{\omega}{kc}\delta b(\mathbf{e}_{x}\pm
i\mathbf{e}_{y})$.
We neglected $\mathbf{E}$ in the right hand side of
Eq.~(\ref{kinEqFurCr}) assuming $v_{\rm{ph}}/v_{s} <<1$. Then using
$\jzCR =e \nCR v_{s}$ from Eq.~(\ref{distrFuncCr}), an integration
over time of Eq.~(\ref{CurrentCr1}) yields
%
%
\begin{equation}\label{deltaJcrNu}
\delta \jCR = g^{\,\prime}\delta
\mathbf{b}\int_{0}^{\infty}\sigma(p)N(p)p^{2}dp,
\end{equation}
\begin{eqnarray}\label{sigmaCrNu}
& &
\sigma=\frac{3}{2x^{2}}+
\frac{3}{8x}\left(1-\frac{1}{x^{2}}+
\left(\frac{a}{x}\right)^{2}\right)\Psi_{1}
-\frac{3a}{2x^{3}}\Psi_{2}\mp
 \nonumber \\
 &\mp& i\left\{\frac{3}{4x}\left(1-\frac{1}{x^{2}}+
\left(\frac{a}{x}\right)^{2}\right)\Psi_{2}
 -\frac{3a}{2x^{2}}+\frac{3a}{2x^{3}}\Psi_{1}\right\},
  \nonumber \\
  & &
\Psi_{1}(x)=ln\left[\frac{(x+1)^{2}+a^{2}}{(x-1)^{2}+a^{2}}\right],
 \nonumber \\
 & &
\Psi_{2}(x)=arctg\left(\frac{x+1}{a}\right)+arctg\left(\frac{x-1}{a}\right).
\end{eqnarray}

\begin{figure}
\centering {
 \rotatebox{0}{
{\includegraphics[height=7.5cm, width=9.5cm]{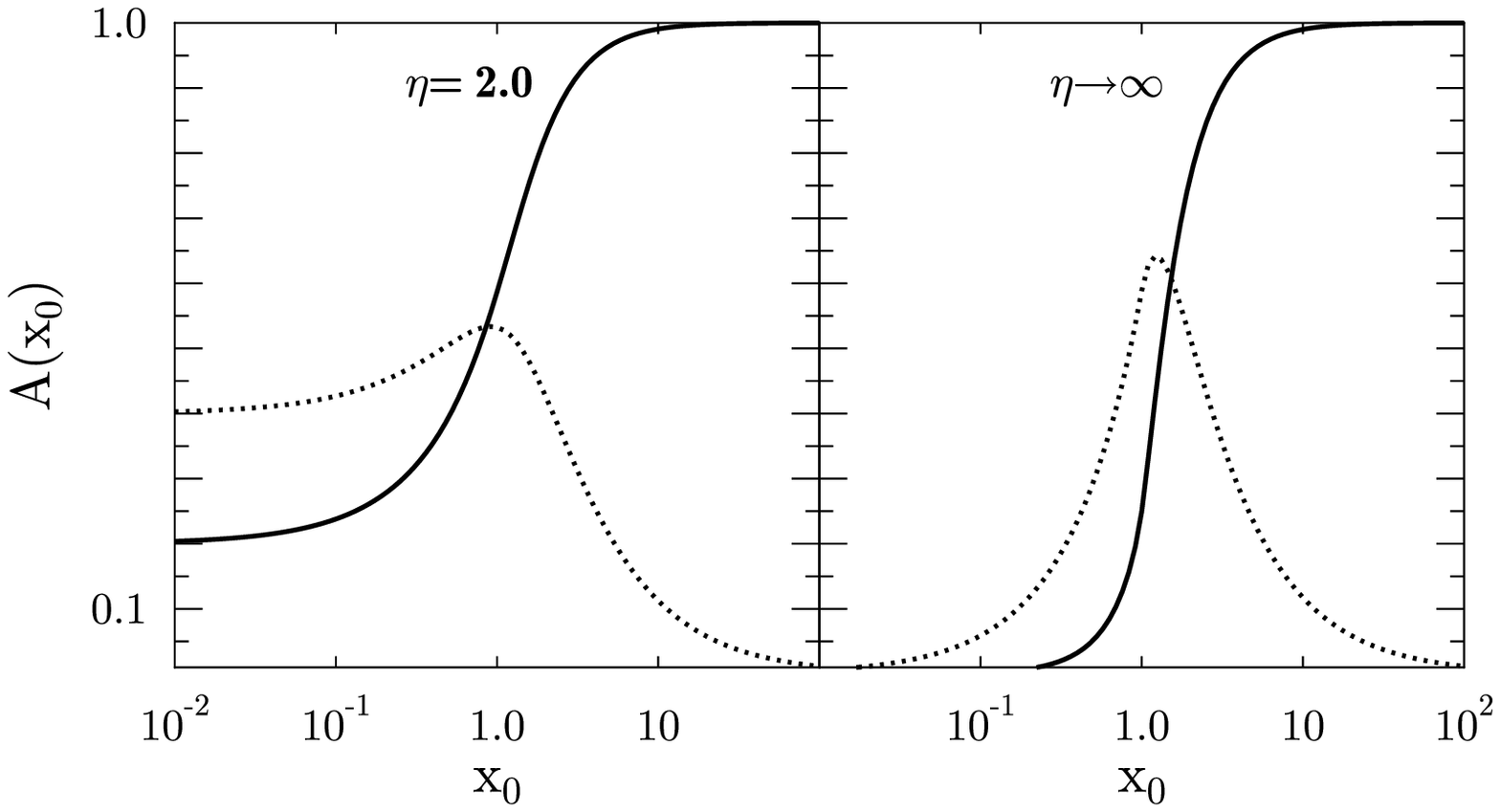} }}}
\caption{The dependence of the dimensionless complex response
function $A(x_{0})$ on $x_{0}=kr_{g0}$. The imaginary parts $\Im
A(x_{0})$ are shown as the dotted curves, while  $1-\Re A(x_{0})$
are the solid curves. Left panel: the \hydro\ case with the
dimensionless particle mean free path $\eta=2$ Right panel: the
limit of Bell (2004) corresponding to $\eta \rightarrow
\infty$.}\label{ResonA}
\end{figure}

For convenience, we introduce the function $A(x_{0})$ defined as
%
\begin{equation}\label{coeffANu}
A(x_{0})=\int_{0}^{\infty}\sigma(p)N(p)p^{2}dp
\ ,
\end{equation}
where $x_{0}=\displaystyle\frac{kcp_{0}}{eB_{0}}$ giving:
\begin{equation}\label{deltaJcrNu1}
\delta \jCR =g^{\,\prime}A(x_{0})\delta\mathbf{b}
\ .
\end{equation}
Then, in the long-wavelength limit $x_{0}\ll 1$, $A(x_{0})$ takes the
form
\begin{equation}\label{coeffANu1}
A(x_{0})\approx
\frac{1}{1 + 1/\eta^{2}} \left(1\mp\frac{i}{\eta}\right).
\end{equation}
In the intermediate limit, $\eta\rightarrow \infty$,
Eq.~(\ref{sigmaCrNu}) has a form equivalent to that of \citet{bell04},
i.e.,
\begin{eqnarray}\label{sigmaCr}
& &
\sigma=\frac{3}{2x^{2}}+\frac{3}{4x}\left(1-\frac{1}{x^{2}}\right)
ln\left|\frac{x+1}{x-1}\right| \mp
\nonumber\\
& & \mp
i\pi\frac{3}{4x}\left(1-\frac{1}{x^{2}}\right)\Theta\left(|x|-1\right),
\end{eqnarray}
where $\Theta(x)$ is the Heaviside function.
For the important case
of a test particle distribution of shock accelerated CRs with
$\alpha=4$,
Eq.~(\ref{coeffANu}) reduces to that of \citet{bell04} and then
\begin{eqnarray}\label{Acoeff}
& &
A(x_{0}) =
\int_{p_{0}}^{\infty}\sigma(p)\frac{p_{0}}{p^{2}}dp =
\nonumber\\
& &
=\frac{3}{8}\left(1+\frac{1}{x_{0}^{2}}\right)-
 \frac{3}{16}x_{0}\left(\frac{1}{x_{0}^{2}}-1\right)^{2}
ln\left|\frac{x_{0}+1}{x_{0}-1}\right|
\mp
\nonumber\\
 & &
 \mp\frac{3\pi i}{16}x_{0}
 \left\{%
\begin{array}{ll}
    \displaystyle\frac{1}{x_{0}^{2}}\left(2-\frac{1}{x_{0}^{2}}\right),
\ \mathrm{for} \ x_{0}>1\\
    \displaystyle 1, \ \mathrm{for} \ x_{0}<1 
\end{array}%
\right ..
\end{eqnarray}
In the short-wavelength limit, $x_{0}\gg 1$  and $A(x_{0})$ has the same
asymptotic behavior as for finite $\eta$ as is clearly seen in
Fig.~(\ref{ResonA}).

\subsection{Long-wavelength CR current and pressure
responses to oblique magnetic perturbations}
In the case of parallel propagating perturbations, the simple
approximations for the linear response of the CR current were obtained
in both the intermediate and the hydrodynamical regimes.
In the case of oblique perturbations, the general equations were
obtained in the form of infinite sequences of Bessel functions with
cyclotron and Cherenkov resonant denominators \citep[see
e.g.,][]{kt73}. Our main aim in this paper is to address the
long-wavelength dynamics of the system with cosmic rays.
If the wavelength exceeds the particle mean free path (the
long-wavelength limit), or in the other words if $kc/\nu$ is small, one
can get simple analytic expressions for both the cosmic-ray current and
the cosmic-ray pressure responses that can be used to obtain the wave
dispersion equations. Note here that, contrary to longitudinal
propagating perturbations, the oblique perturbations induce a cosmic-ray
pressure response.

The linear response function satisfies
\begin{eqnarray}\label{KpertCR}
& & \frac{\partial \delta f}{\partial
t}+\mathbf{v}\cdot\frac{\partial \delta
f}{\partial\mathbf{r}}-\frac{ec}{\mathcal{E}}\mathbf{B}_0
\cdot\mathbf{\mathcal{O}}\delta f=
\nonumber \\
  & &
=-e\mathbf{E}\cdot\frac{\partial
f_0}{\partial\mathbf{p}}+\frac{ec}{\mathcal{E}}\delta\mathbf{b}
\cdot\mathbf{\mathcal{O}}f_0-\nu(\delta f - \Delfiso).
\end{eqnarray}
In the long-wavelength limit, one can neglect terms of the order
of  $kc/\nu \ll 1$.
%
Now, we define a coordinate system with the $z$-axis
along the unperturbed initial magnetic field, $\mathbf{B_{0}}$,
the polar angle $\theta$, and the azimuthal angle
$\varphi$ in the $x-y$ plane.
Thus, $k_{\parallel} = k \cos\theta$
and $k_{\perp} = k\sin\theta$.
To obtain a solution to Eq.~(\ref{KpertCR}), we neglect the electric
field $\mathbf{E}$ on the right-hand-side, assuming $v_{\rm{ph}} \ll
v_{s}$, and also assume the low frequency perturbation $\omega \ll
ck$.
Then, the solution to Eq.~(\ref{KpertCR}) can be presented as a
superposition of spherical harmonics in $\theta$ and $\varphi$ with
coefficients that depend on the particle energy.
\begin{eqnarray}\label{KleftPart}
& & \frac{ec}{\mathcal{E}}\delta\mathbf{b}
\cdot\mathbf{\mathcal{O}}f_0 = S_{1}\sin{\theta} \cos{\varphi} +
S_{2} \sin{\theta} \sin{\varphi},
\nonumber \\
& & S_{1}=-\frac{e}{p}\delta b_{y}\frac{3v_{s}}{c}\frac{\nCR
N(p)}{4\pi},
\nonumber \\
& &
 S_{2}=\frac{e}{p}\delta b_{x}\frac{3v_{s}}{c}\frac{\nCR
N(p)}{4\pi} \ .
\end{eqnarray}
Then, presenting the angular dependence of the distribution function
as
\begin{eqnarray}\label{fucDistCRp}
& & \delta f(\mathbf{p}) = A_{0}(p)+A_{1}(p) \cos{\theta} +
\nonumber \\
& &
 +A_{2}(p)
\sin{\theta} \cos{\varphi} + A_{3}(p) \sin{\theta} \sin{\varphi},
\end{eqnarray}
 one obtains, in the limit $ck/\nu \ll 1$, the four equations for the
 coefficients:
%
$$%
ik_{\parallel}cA_{0}(p)+\frac{1}{\eta}\Omega A_{1}(p)=0,$$
$$%
ik_{x}cA_{0}(p)+\frac{1}{\eta}\Omega A_{2}(p)-\Omega
A_{3}(p)=S_{1},$$
$$%
ik_{y}cA_{0}(p)+\Omega A_{2}(p)+\frac{1}{\eta}\Omega
A_{3}(p)=S_{2},$$
\begin{equation}\label{KsystEq}
k_{\parallel}A_{1}(p)+k_{x}A_{2}(p)+k_{y}A_{3}(p)=0.
\end{equation}
%
Using the above, the linear responses of the cosmic-ray current and
pressure can be obtained from
%
%
$$%
\delta \jCR =e\int \mathbf{v}\delta f(\mathbf{p})d^{3}p,
$$
\begin{equation}\label{curPressCR}
\delta \PCR =\frac{1}{3}\int vp\delta f(\mathbf{p})d^{3}p.
\end{equation}

Substituting the solutions of Eqs.~(\ref{KsystEq}), into
(\ref{fucDistCRp}) and (\ref{curPressCR}), and performing the
integration over the particle momentum $d^{3}p$, we  obtain finally
the cosmic-ray current and pressure response on the imposed oblique
magnetic fluctuations:
$$%
\delta \mathbf{j}^{cr} =
\frac{4\pi}{3} ec\int^{\infty}_{0}
\left\{A_{1}(p)\mathbf{e}_{\parallel}+A_{2}(p)\mathbf{e}_{x}+A_{3}(p)\mathbf{e}_{y}\right\}p^{2}dp=
$$
$$%
=g^{\,\prime}\frac{k_{\parallel}}{k_{\parallel}^{2}+\frac{1}{\eta^{2}}k^{2}}
\left\{k_{\parallel}\delta
\mathbf{b}+\frac{1}{\eta}\mathbf{k}\times\delta \mathbf{b}\right\},
$$
\begin{eqnarray}\label{curCRxy}
& &
\delta P_{cr}=\frac{4\pi}{3} c\int^{\infty}_{0} A_{0}(p)p^{3}dp=
\nonumber \\
  & &
=\frac{iB_{0}}{c}g^{\,\prime}
\frac{\mathbf{e}_{\parallel}}{k_{\parallel}^{2}+\frac{1}{\eta^{2}}k^{2}}
\frac{1}{\eta}\left\{k_{\parallel}\delta
\mathbf{b}+\frac{1}{\eta}\mathbf{k}\times\delta \mathbf{b}\right\}.
\end{eqnarray}

We note that in the case of the short-wavelength limit,
 the procedure described
above cannot be applied because
$\displaystyle\frac{kc}{\nu}>1$. In this case, one should treat
cyclotron and Cherenkov resonances.

\section{Correlations in Bell modes}\label{corr}
To obtain the mean field dynamic equations in a closed form one must
express the mean electromotive force
$\mathbf{\overline{\mathcal{E}}}$,
$\left\langle(\mathbf{v}\nabla)\mathbf{v}\right\rangle$,
$\left\langle(\mathbf{b}\nabla)\mathbf{b}\right\rangle$, and other terms
through the correlators of the short-scale Bell fluctuations
 $\left\langle
v_{\alpha}^{*}(\mathbf{k})b_{\beta}(\mathbf{k})\right\rangle$ and
$\left\langle
b_{\alpha}^{*}(\mathbf{k})b_{\beta}(\mathbf{k})\right\rangle$. Since
the maximal growth rates of the Bell modes are along the local mean
magnetic field, it is convenient to present the correlations in a
coordinate system with the $z$-axis along $\mathbf{e}_{z'} =
\mathbf{\overline{B}}/\mid\mathbf{\overline{B}}\mid$,
 where $\mathbf{e}_{x',y'}$ are the unit vectors in the plane transverse
to $\mathbf{e}_{z'}$.
Doing this we obtain:
\begin{eqnarray}\label{Correlbb}
& & \left\langle
b_{\alpha}^{*}(\mathbf{k})b_{\beta}(\mathbf{k})\right\rangle
=\frac{\left\langle\mathbf{b}^{2}(k_{z'})\right\rangle}{2}\times
\nonumber \\
  & &
\times\delta(k_{x'})\delta(k_{y'})\left(%
\begin{array}{ccc}
  1 & -i\displaystyle\frac{k_{z'}}{|k_{z'}|} & 0 \\
  i\displaystyle\frac{k_{z'}}{|k_{z'}|} & 1 & 0 \\
  0 & 0 & 0 \\
\end{array}%
\right),
\end{eqnarray}
\begin{equation}\label{Correlvv}
\left\langle
v_{\alpha}^{*}(\mathbf{k})v_{\beta}(\mathbf{k})\right\rangle
=\frac{1}{4\pi\rho}\frac{k_{1}}{|k_{z'}|}\left\langle
b_{\alpha}^{*}(\mathbf{k})b_{\beta}(\mathbf{k})\right\rangle
\end{equation}
%
\begin{eqnarray}\label{Correlvb1}
& & \left\langle
v_{\alpha}^{*}(\mathbf{k})b_{\beta}(\mathbf{k})\right\rangle
=\frac{1}{\sqrt{4\pi\rho}}\sqrt{\frac{k_{1}}{|k_{z'}|}}\frac{\left\langle\mathbf{b}^{2}(k_{z'})\right\rangle}{2}\times
\nonumber \\
  & &
\times\delta(k_{x'})\delta(k_{y'})\left(%
\begin{array}{ccc}
  i\displaystyle\frac{k_{z'}}{|k_{z'}|} & 1 & 0 \\
  -1 & i\displaystyle\frac{k_{z'}}{|k_{z'}|} & 0 \\
  0 & 0 & 0 \\
\end{array}%
\right),
\end{eqnarray}
\begin{equation}\label{Correlvb}
\left\langle
b_{\alpha}^{*}(\mathbf{k})v_{\beta}(\mathbf{k})\right\rangle
=-\left\langle
v_{\alpha}^{*}(\mathbf{k})b_{\beta}(\mathbf{k})\right\rangle.
\end{equation}

\section{Mean-field induction equation}\label{MFE}
To get the mean field equations we apply an averaging procedure that is
widely used in the mean field dynamo theory \citep{bf02, bs05,
brandenburg09}. The distinctive feature of our model is the presence of
the cosmic-ray current and we shall derive the cosmic-ray current effect
on the mean field dynamics and unstable modes. The electromotive force,
$\mathbf{\overline{\mathcal{E}}}$, satisfies the equation
%
%
\begin{equation}\label{ElectrMot1}
c\frac{\partial\mathbf{\overline{\mathcal{E}}}}{\partial
t}=\left\langle\mathbf{\frac{\partial v}{\partial t}}\times
\mathbf{b}\right\rangle +\left\langle \mathbf{v}\times
\mathbf{\frac{\partial b}{\partial t}}\right\rangle.
\end{equation}

To treat $\partial \Eover/\partial t$
in Eq.~(\ref{ElectrMot1}) we obtain from
Eq.~(\ref{eqMotiontot1}) the equation for short-scale variations of
the bulk velocity, $\mathbf{v}$, including second-order correlations:
\begin{eqnarray}\label{eqMotion}
 & &\frac{\partial \mathbf{v}}{\partial t} =
   -\frac{1}{c\rho}
((\mathbf{\jCRmean} -e \nCR \mathbf{\overline{V}})\times\mathbf{b})
       +
     \nonumber \\
  & &
  +\frac{e \nCR}{c\rho}(\mathbf{v}\times\mathbf{\overline{B}})
  -(\mathbf{\overline{V}}\nabla)\mathbf{v}-(\mathbf{v}\nabla)\mathbf{\overline{V}}+
  \nonumber \\
  & &
+\frac{1}{4\pi\rho}(\nabla\times\mathbf{b})\times\mathbf{\overline{B}}+
     \frac{1}{4\pi\rho}(\nabla\times\mathbf{\overline{B}})\times\mathbf{b}+
  \nonumber \\
    & &
     +\frac{1}{4\pi\rho}(\nabla\times\mathbf{b})\times\mathbf{b}-
     \frac{1}{4\pi\rho}
\left\langle(\nabla\times\mathbf{b})\times\mathbf{b}\right\rangle-
      \nonumber \\
  & &
  -(\mathbf{v}\nabla)\mathbf{v} +
\left\langle(\mathbf{v}\nabla)\mathbf{v}\right\rangle
  +\frac{e \nCR}{c\rho}(\mathbf{v}\times\mathbf{b})-
      \nonumber \\
  & &
  -\frac{e \nCR}{c\rho}\left\langle\mathbf{v}\times\mathbf{b}\right\rangle
  +\nu\triangle \mathbf{v},
\end{eqnarray}
%
%
%
where $\jCRmeanVec$
is the averaged cosmic-ray current. In Eq.~(\ref{eqMotion}), the
short-scale density fluctuations were omitted because the fastest
growing modes are the incompressible modes with wave vectors along the
local mean magnetic field. The short-scale fluctuations of the
cosmic-ray current were also neglected being small for
$kcp_0/(eB_0) > 1$
as shown in Appendix~\ref{CRCR}.

The fluctuating part of the magnetic field, $\mathbf{b}$, satisfies
the equation
\begin{eqnarray}\label{eqInduct}
& & \frac{\partial\mathbf{b}}{\partial
t}=\nabla\times(\mathbf{v}\times\mathbf{\overline{B}})
+\nabla\times(\mathbf{\overline{V}}\times\mathbf{b})+
\nonumber \\
& &
+\nabla\times(\mathbf{v}\times\mathbf{b})-\nabla\times\left\langle\mathbf{v}\times\mathbf{b}\right\rangle
+\nu_{m}\triangle\mathbf{b}.
\end{eqnarray}

Substituting Eqs.~(\ref{eqMotion}) and (\ref{eqInduct}) into
Eqs.~(\ref{ElectrMot1}), (\ref{VnablaV}), and (\ref{BnablaB}), one
obtains:
%
%
\begin{eqnarray}\label{ElectrMot11}
  & & c\frac{\partial\mathbf{\overline{\mathcal{E}}}}{\partial
t} = -\frac{1}{c\rho}
\left\langle((\mathbf{\jCRmean} -e \nCR\mathbf{\overline{V}})
\times\mathbf{b})\times\mathbf{b}\right\rangle+
\nonumber \\
    & &
+\left\langle\mathbf{v}\times\{\nabla\times
(\mathbf{v}\times\mathbf{\overline{B}})+\nabla\times
(\mathbf{\overline{V}}\times\mathbf{b})\}\right\rangle-
     \nonumber \\
    & &
    -\left\langle\{(\mathbf{\overline{V}}\nabla)\mathbf{v} +
(\mathbf{v}\nabla)\mathbf{\overline{V}}\}\times\mathbf{b}\right\rangle+
    \nonumber \\
    & &
    +\frac{1}{4\pi\rho}\left\langle\{(\nabla\times\mathbf{b})
\times\mathbf{\overline{B}})+((\nabla\times\mathbf{\overline{B}})
\times\mathbf{b})\}\times\mathbf{b}\right\rangle+
  \nonumber \\
  & &
  +\frac{e \nCR}{c\rho}\left\langle(\mathbf{v}\times
\mathbf{\overline{B}})\times\mathbf{b}\right\rangle
  +\nu\left\langle\triangle \mathbf{v}\times\mathbf{b}\right\rangle+
    \nonumber \\
  & &
  +\nu_{m}\left\langle\mathbf{v}\times\triangle\mathbf{b}\right\rangle - c
\frac{\mathbf{\overline{\mathcal{E}}}}{\taucor}
\ .
\end{eqnarray}
The last term, $\Eover/\taucor$,
in Eq.~(\ref{ElectrMot11}) approximates the time relaxation of
triple correlations with the time scale $\taucor$ \citep[see,
e.g.,][]{bs05}.
The correlation time is typically expected to be about the turnover time
of the turbulence. For the sake of simplicity we use the same relaxation
time $\taucor$ for all of the triple correlations.  A more rigorous
analysis that distinguishes the correlation times of different triple
correlations is beyond of the scope of this paper and will be done
separately. The time derivative, $\partial \Eover/\partial t$, in
Eq.~(\ref{ElectrMot11}) suppresses the mean-field variations on time
scales below $\taucor$ and is analogous to the Faraday displacement
current in Maxwell's equations \citep[e.g.,][]{brandenburg09}.  This
term can be omitted in the analysis of the long-wavelength modes of
frequencies $\omega \taucor \ll$ 1 in the close analogy with the
well-known MHD approximation.

For Bell's instability, the
terms in Eq.~(\ref{ElectrMot11}) containing the short-scale
magnetic field correlators Eq.~(\ref{Correlbb}) and
Eq.~(\ref{Correlvb}) (apart from a term with the cosmic-ray current)
are $k_{1}/|k_{z'}| \gg 1$ times smaller than the velocity
correlator Eq.~(\ref{Correlvv}). With this approximation,  the
equation for the mean electromotive field simplifies to
%
\begin{eqnarray}\label{ElectrMot2}
  & & c\frac{\partial\mathbf{\overline{\mathcal{E}}}}{\partial
t}=\frac{1}{\rho
c}\left\langle\mathbf{b}^{2}
\right\rangle
\left(\frac{1}{2}(\jCRx -e \nCR \overline{V}_{x'})\mathbf{e}_{x'} +
\right .
  \nonumber \\
  & &
\left .
+\frac{1}{2}(\jCRy -e \nCR \overline{V}_{y'})\mathbf{e}_{y'}
+(\jCRz -e \nCR\overline{V}_{z'})\mathbf{e}_{z'}\right)-
     \nonumber \\
    & &
-\left\langle\mathbf{v}\cdot\nabla\times
\mathbf{v}\right\rangle\overline{B}_{z'}\mathbf{e}_{z'}-\frac{\left\langle
v^{2}\right\rangle}{2}\nabla_{\perp}\times \overline{\mathbf{B}} - c
\frac{\mathbf{\overline{\mathcal{E}}}}{\taucor}
\end{eqnarray}
where $\nabla_{\perp}$ contains transverse coordinate derivatives
only. 

Using Eq.~(\ref{Correlvv}), the second term on the
right-hand-side of Eq.~(\ref{ElectrMot2}) can be written as
%
\begin{equation}\label{sotnJzB}
-\left\langle\mathbf{v}\cdot\nabla\times\mathbf{v}\right\rangle
\overline{B}_{z'}\mathbf{e}_{z'}=
\frac{1}{\rho
c}\left\langle\mathbf{b}^{2}\right\rangle
\left(\jCRz -e \nCR \overline{V}_{z'}\right)\mathbf{e}_{z'}
\ .
\end{equation}
Then, the mean field induction equation yields
\begin{eqnarray}\label{MediumInd1}
  & & \frac{\partial\mathbf{\overline{B}}}{\partial
t}=\nabla\times(\mathbf{\overline{V}}\times\mathbf{\overline{B}})+
  \nonumber \\
    & &
    +\frac{2}{\rho
c}\left\langle\mathbf{b}^{2}\right\rangle\taucor \nabla\times
\left(\jCRz -e \nCR \overline{V}_{z'}\right)\mathbf{e}_{z'}+
     \nonumber \\
    & & +\frac{1}{2\rho
c}\left\langle\mathbf{b}^{2}\right\rangle\taucor \nabla\times
\left(\left(\jCRx -e \nCR \overline{V}_{x'}\right)\mathbf{e}_{x'}
+ \right .
  \nonumber \\
    & & \left .
+\left(\jCRy -e \nCR \overline{V}_{y'}\right)
\mathbf{e}_{y'}\right)-\frac{\left\langle
v^{2}\right\rangle}{2}\taucor\nabla\times\left(\nabla_{\perp}\times
\overline{\mathbf{B}}\right) .
\end{eqnarray}

The long-wavelength perturbations $\delta \jCRVec$
 in Eqs.~(\ref{ElectrMot2}) and  (\ref{largeEqMotion0}) were derived in
a coordinate system with the axis $\mathbf{e}_{z'} =
\mathbf{\overline{B}}/\mid\mathbf{\overline{B}}\mid$ along the local
mean magnetic field. To make a linear analysis of the dispersion
relations it is convenient to transform the vector coordinates to
the laboratory system where the $z$-axis is along the unperturbed
magnetic field and the shock normal $\mathbf{e}_{z} =
\mathbf{{B_0}}/{B}_0$.
To first order in the small amplitudes of
the current and field perturbations, the transformation yields:
\begin{equation}\label{preobr1Ind}
(\jCRx -e \nCR \overline{V}_{x'})\mathbf{e}_{x'}
\approx[-g^{\,\prime}\delta
b_{x} +
(\djCRx - e \nCR \delta\overline{V}_{x})]\mathbf{e}_{x},
\end{equation}
\begin{equation}\label{preobr2Ind}
(\jCRy -e \nCR \overline{V}_{y'})\mathbf{e}_{y'}
\approx[-g^{\,\prime}\delta
b_{y}+(\djCRy - e \nCR \delta\overline{V}_{y})]\mathbf{e}_{y}
\ ,
\end{equation}
\begin{equation}\label{preobr3Ind}
(\overline{j}^{cr}_{z'}-en_{cr}\overline{V}_{z'})\mathbf{e}_{z'}\approx(
\overline{j}^{cr}_{z}-en_{cr}\overline{V}_{z})\mathbf{e}_{z}+g^{\,\prime}\delta
b_{x}\mathbf{e}_{x}+g^{\,\prime}\delta b_{y}\mathbf{e}_{y} \ .
\end{equation}

\section{The averaged momentum equation}\label{EMotion}
To get the equation of motion averaged over the short-scale
fluctuations in a closed form, we derive
$\left\langle(\mathbf{v}\nabla)\mathbf{v}\right\rangle$ and
$\left\langle(\mathbf{b}\nabla)\mathbf{b}\right\rangle$ in
Eq.~(\ref{largeEqMotion}) for large-scale motions using the mean-field
approximation as it was described in Appendix~\ref{MFE}. This yields:

\begin{equation}\label{VnablaV}
\frac{\partial }{\partial
t}\left\langle(\mathbf{v}\nabla)\mathbf{v}\right\rangle=
\left\langle\left(\frac{\partial\mathbf{v}}{\partial
t}\nabla\right)\mathbf{v}\right\rangle+\left\langle\left(\mathbf{v}\nabla\right)\frac{\partial\mathbf{v}}{\partial
t}\right\rangle,
\end{equation}
\begin{equation}\label{BnablaB}
\frac{\partial }{\partial
t}\left\langle(\mathbf{b}\nabla)\mathbf{b}\right\rangle=\left\langle\left(\frac{\partial\mathbf{b}}{\partial
t}\nabla\right)\mathbf{b}\right\rangle+\left\langle\left(\mathbf{b}\nabla\right)\frac{\partial\mathbf{b}}{\partial
t}\right\rangle,
\end{equation}

\begin{eqnarray}\label{VnablaV1}
  & & \frac{\partial }{\partial
t}\left\langle(\mathbf{v}\nabla)\mathbf{v}\right\rangle +
\frac{1}{\taucor}\left\langle(\mathbf{v}\nabla)\mathbf{v}\right\rangle=
   \nonumber \\
 & &
  =-\frac{1}{c\rho}\left\langle(((\jCRmeanVec
  -e\nCR\mathbf{\overline{V}})
\times\mathbf{b})\nabla)\mathbf{v}\right\rangle-
     \nonumber \\
 & &
-\frac{1}{c\rho}\left\langle(\mathbf{v}\nabla)((\jCRmeanVec -e\nCR
\mathbf{\overline{V}})\times\mathbf{b})\right\rangle-
   \nonumber \\
 & &
-\left\langle(\{(\mathbf{\overline{V}}\nabla)\mathbf{v}+(\mathbf{v}\nabla)
\mathbf{\overline{V}}\}\nabla)\mathbf{v}\right\rangle-
      \nonumber \\
    & &
    -\left\langle(\mathbf{v}\nabla)\{(\mathbf{\overline{V}}\nabla)\mathbf{v}+(\mathbf{v}\nabla)\mathbf{\overline{V}}\})\right\rangle+
       \nonumber \\
 & &
   + \frac{1}{4\pi\rho}\left\langle(\mathbf{v}\nabla)\{(\nabla\times\mathbf{b})\times\mathbf{\overline{B}})+((\nabla\times\mathbf{\overline{B}})\times\mathbf{b})\}\right\rangle+
  \nonumber \\
  & &
  +\frac{1}{4\pi\rho}\left\langle(\{(\nabla\times\mathbf{b})\times\mathbf{\overline{B}}+(\nabla\times\mathbf{\overline{B}})\times\mathbf{b}\}\nabla)\mathbf{v}\right\rangle+
    \nonumber \\
 & &
+\frac{e\nCR}{c\rho}
\left\langle((\mathbf{v}\times\mathbf{\overline{B}})\nabla)\mathbf{v}\right\rangle
     +\frac{e\nCR}{c\rho}
\left\langle(\mathbf{v}\nabla)(\mathbf{v}\times\mathbf{\overline{B}})\right\rangle+
         \nonumber \\
 & &
     +\nu\left\langle(\triangle \mathbf{v}\nabla)\mathbf{v}\right\rangle+
\nu\left\langle(\mathbf{v}\nabla)\triangle \mathbf{v}\right\rangle,
\end{eqnarray}

\begin{eqnarray}\label{BnablaB1}
  & & \frac{\partial }{\partial
t}\left\langle(\mathbf{b}\nabla)\mathbf{b}\right\rangle+
\frac{1}{\taucor}\left\langle(\mathbf{b}\nabla)\mathbf{b}\right\rangle=
    \nonumber \\
 & &
  =\left\langle(\{\nabla\times(\mathbf{v}\times\mathbf{\overline{B}})+\nabla\times(\mathbf{\overline{V}}\times\mathbf{b})\}\nabla)\mathbf{b}\right\rangle+
     \nonumber \\
    & &
+\left\langle(\mathbf{b}\nabla)\{\nabla\times(\mathbf{v}\times\mathbf{\overline{B}})+\nabla\times(\mathbf{\overline{V}}\times\mathbf{b})\}\right\rangle+
    \nonumber \\
    & &
    +\nu_{m}\left\langle(\triangle
\mathbf{b}\nabla)\mathbf{b}\right\rangle
    +\nu_{m} \left\langle(\mathbf{b}\nabla)\triangle
    \mathbf{b}\right\rangle.
\end{eqnarray}

Then, the equation of motion in the mean field approximation is
\begin{eqnarray}\label{largeEqMotion0}
  & & \frac{ \partial\mathbf{\overline{V}}}{\partial t }
   +(\mathbf{\overline{V}}\nabla)\mathbf{\overline{V}}
   =  -\frac{1}{\rho}\nabla P-
\nonumber \\
& &
    -\frac{1}{c\,\rho}(
(\jCRmeanVec -e \nCR\mathbf{\overline{V}})\times\mathbf{\overline{B}})
    +\frac{1}{4\pi\rho}((\nabla\times\mathbf{\overline{B}})\times\mathbf{\overline{B}})
    + \nonumber \\
    & &
    +\frac{\kappa_{t}}{c\rho}((\jCRx
   -e\nCR\overline{V}_{x'})\mathbf{e}_{y'}-
(\jCRy -e\nCR \overline{V}_{y'})\mathbf{e}_{x'})+
\nonumber \\
& & +\frac{\zeta_{t}}{c\rho} \left(\frac{\partial(\jCRx
-e\nCR\overline{V}_{x'})}{\partial
x'}\mathbf{e}_{z'}+\frac{\partial(\jCRy
-e\nCR\overline{V}_{y'})}{\partial y'}\mathbf{e}_{z'}- \right.
\nonumber \\
& & -\left.
         \frac{\partial(\jCRz -e\nCR \overline{V}_{z'})}{\partial x'}\mathbf{e}_{x'}
            -\frac{\partial(\jCRz -e\nCR \overline{V}_{z'})}{\partial y'}\mathbf{e}_{y'}
       \right)+
       \nonumber \\
   & &
       +\taucor\left\langle\mathbf{v}\cdot\nabla\times\mathbf{v}\right\rangle
\left(\frac{\partial{\overline{V}}_{z'}}{\partial y'}\mathbf{e}_{x'}
-\frac{\partial{\overline{V}}_{z'}}{\partial
x'}\mathbf{e}_{y'}\right)+
\nonumber \\
   & &
+\taucor\frac{\left\langle v^{2}\right\rangle}{2}
      \left(\frac{\partial^{2}\mathbf{\overline{V}}}{\partial x'^{2}}
           +\frac{\partial^{2}\mathbf{\overline{V}}}{\partial y'^{2}}
      \right)+\nu\triangle\mathbf{\overline{V}}-
  \nonumber \\
  & &
-\frac{1}{c\,\rho}\left\langle(\jCRVec -e\nCR\mathbf{v})\times\mathbf{b}\right\rangle...,
\end{eqnarray}
where the turbulent transport coefficients
\begin{equation}\label{turbcoef1}
\kappa_{t} = \taucor \int^{\infty}_{0} \!\! dk_{z'}
              \frac{1}{\sqrt{4\pi\rho}}\sqrt{k_{1}|k_{z'}|}
              \left\langle \mathbf{b}^{2}(k_{z'})\right\rangle
\ ,
\end{equation}
and
%
%
\begin{equation}\label{turbcoef2}
\zeta_{t} = \taucor \int^{\infty}_{0}   \!\! dk_{z'}
              \frac{1}{\sqrt{4\pi\rho}}\sqrt{\frac{k_{1}}{|k_{z'}|}}
              \left\langle \mathbf{b}^{2}(k_{z'})\right\rangle
\end{equation}
%
are expressed through the  correlation time $\taucor$.

As in Appendix~\ref{MFE}, we transform the vector coordinates to the
 laboratory system where the $z$-axis is along the unperturbed
magnetic field and the shock normal $\mathbf{e}_{z} =
\mathbf{{B_0}}/{B}_0$. Then, to first-order in the small
amplitudes of
the current and field perturbations the transformation yields
\begin{equation}\label{preobr1}
(\jCRx -e\nCR \overline{V}_{x'})\mathbf{e}_{y'}\approx
[-g^{\,\prime}\delta
b_{x}+(\djCRx - e\nCR\delta\overline{V}_{x})]\mathbf{e}_{y},
\end{equation}
\begin{equation}\label{preobr2}
(\jCRy - e\nCR \overline{V}_{y'})\mathbf{e}_{x'}\approx
[-g^{\,\prime}\delta
b_{y}+(\djCRy - e\nCR \delta\overline{V}_{y})]\mathbf{e}_{x}.
\end{equation}
%

\section*{Acknowledgments}
We thank the anonymous referee for careful reading of our paper and
useful comments. Some of the calculations were performed at the Joint
Supercomputing Centre (JSCC RAS) and the Supercomputing Centre at Ioffe
Institute, St.Petersburg. A.M.B. and S.M.O. were supported in part by
RBRF grant 09-02-12080 and by the RAS Presidium Programm. D.C.E
acknowledges support from NASA grants ATP02-0042-0006,
NNH04Zss001N-LTSA, and 06-ATP06-21.  A.M.B. and D.C.E. gratefully
acknowledge the KITP program in Santa Barbara ``Particle Acceleration in
Astrophysical Plasmas.''

\end{document}